\documentclass[a4paper,11pt]{article}
\pdfoutput=1 

\usepackage{jheppub} 

\usepackage{amsmath,amssymb,bm,bbm}
\usepackage[utf8]{inputenc}

\title{\boldmath Rectangular W-algebras of types {$so(M)$} and {$sp(2M)$} and dual coset CFTs}

\author[a]{Thomas Creutzig,}
\author[b]{Yasuaki Hikida}
\author[c]{and Takahiro Uetoko}

\affiliation[a]{Department of Mathematical and Statistical Sciences, University of Alberta,\\Edmonton, Alberta T6G 2G1, Canada}

\affiliation[b]{Center for Gravitational Physics, Yukawa Institute for Theoretical Physics, Kyoto University,\\Kyoto 606-8502, Japan}

\affiliation[c]{Department of Physical Sciences, College of Science and Engineering, Ritsumeikan University,\\ Shiga 525-8577, Japan}

\emailAdd{creutzig@ualberta.ca}
\emailAdd{yhikida@yukawa.kyoto-u.ac.jp}
\emailAdd{rp0019fr@ed.ritsumei.ac.jp}

\abstract{We examine rectangular W-algebras with $so(M)$ or $sp(2M)$ symmetry, which can be realized as the asymptotic symmetry of higher spin gravities with restricted matrix extensions. We compute the central charges of the algebras and the levels of  $so(M)$ or $sp(2M)$ affine subalgebras by applying the Hamiltonian reductions of $so$ or $sp$ type Lie algebras. For simple cases with generators of spin up to two, we obtain their operator product expansions by requiring the associativity. We further claim that the W-algebras can be realized as the symmetry algebras of dual coset CFTs and provide several strong supports. The analysis can be regarded as a check of extended higher spin holographies including full quantum corrections. We also extend the analysis by introducing $\mathcal{N}=1$ supersymmetry.
}

\arxivnumber{1906.05872}
\preprint{YITP-19-50}

\begin{document}
	\maketitle
	\flushbottom

\section{Introduction and summary}

It has been argued that higher spin gravity is useful to understand the tensionless limit of superstring theory \cite{Gross:1988ue}. In particular, the matrix extended version of Vasiliev theory is expected to explain  higher Regge trajectories of superstrings, see, e.g., \cite{Vasiliev:2018zer} including higher tensor extensions.
In \cite{Creutzig:2013tja}, it was proposed that classical 3d Prokushkin-Vasiliev theory with $M \times M$ matrix valued fields constructed in \cite{Prokushkin:1998bq} is dual to 2d Grassmannian-like model 
\begin{align}
\frac{su(N+M)_k }{su(N)_k \oplus u(1)_{k N M (N+M)}} \label{coset}
\end{align}
at a large $N$ limit, see also \cite{Candu:2013fta,Eberhardt:2018plx,Kumar:2018dso} for related works.
Setting $M=1$, the proposal essentially reduces to the Gaberdiel-Gopakumar duality \cite{Gaberdiel:2010pz} 
and its $\mathcal{N}=2$ version becomes the same as that of \cite{Creutzig:2011fe}.
With $\mathcal{N}=3$ enhanced  supersymmetry, the relation to superstring theory on AdS$_3 \times $M$_7$ has been discussed \cite{Creutzig:2014ula,Hikida:2015nfa}, where M$_7$ is a 7-dimensional manifold.
It would be useful to introduce $\mathcal{N}=4$ supersymmetry as in \cite{Gaberdiel:2013vva,Gaberdiel:2014cha}, which enables us to relate to superstring theory on AdS$_3 \times $S$^3 \times $S$^3 \times $S$^1$ or AdS$_3 \times $S$^3 \times $T$^4$. However, it seems difficult to introduce the matrix degrees of freedom in a way compatible with the $\mathcal{N}=4$ supersymmetry.

In \cite{Creutzig:2018pts}, the asymptotic symmetry of the extended higher spin gravity was examined including full quantum corrections.%
\footnote{Without the matrix extension, the quantum asymptotic symmetry was examined in \cite{Gaberdiel:2012ku} for the bosonic case and in \cite{Candu:2012tr} for the $\mathcal{N}=2$ supersymmetric case.}
Three dimensional higher spin gravity can be constructed by a Chern-Simons gauge theory based on a higher rank gauge algebra $\mathfrak{g}$.
Without matrix extension, the gauge algebra of the Prokushkin-Vasiliev theory is given by $hs[\lambda]$, which can be truncated to $sl(n)$ at $\lambda = n$ $(n=2,3,\ldots)$. 
Setting $\mathfrak{g} = sl(n)$, the higher spin gravity includes the gauge fields of spin $s=2,3,\ldots,n$.
We can extend  $hs[\lambda]$ by multiplying $M \times M$ matrix degrees of freedom \cite{Gaberdiel:2013vva,Creutzig:2013tja}, and the extended algebra may be denoted as $hs_M[\lambda]$.
The gauge algebra is truncated to $sl(Mn)$ at $\lambda = n$, and it may be decomposed as \cite{Joung:2017hsi,Creutzig:2018pts}
\begin{align}
sl(Mn) \simeq sl(M) \otimes \mathbbm{1}_n \oplus \mathbbm{1}_M \otimes sl(n) \oplus sl(M) \otimes sl(n) \, . \label{sldec}
\end{align}
In particular, the higher spin gravity has $sl(M)$ (or $su(M)$)
gauge sector.%
\footnote{Throughout this paper, we consider the complex elements of gauge algebras, therefore, for instance, we do not distinguish  $sl(M)$ and $su(M)$.}
The gravitational sector is identified with the principally embedded $sl(2)$ in $\mathbbm{1}_M \otimes sl(n)$.
Applying the general prescription of \cite{Henneaux:2010xg,Campoleoni:2010zq}, the asymptotic symmetry can be identified as the rectangular W-algebra with $su(M)$ symmetry.
The central charge $c$ of the algebra and the level $\ell$ of $su(M)$ currents were computed and the operator product expansions (OPEs) among generators of spin up to two were obtained by requiring  associativity of the OPEs. 
Based on the holography of \cite{Creutzig:2013tja}, it was conjectured that the rectangular W-algebra can be realized by the symmetry algebra of \eqref{coset} even with finite $c$, and several important checks have been provided. The analysis of symmetry with finite $c$ would allow us to obtain some information on quantum effects of higher spin gravity.

In this paper, we generalize the analysis by considering the restricted matrix extensions of higher spin gravity. There are two ways to truncate the degrees of freedom \cite{Prokushkin:1998bq}, and the higher spin gravity includes $so(M)$ or $sp(2m)$ (for $M = 2m$) gauge sector depending on the way of truncation.
Without matrix extensions, the gauge algebra $hs[\lambda]$ can be truncated to $hs^\text{e}[\lambda]$ with only even spin generators. The holography with the truncated higher spin gravity was proposed in \cite{Ahn:2011pv,Gaberdiel:2011nt} and the quantum asymptotic symmetry were analyzed in \cite{Candu:2012ne}.
The even spin algebra $hs^\text{e}[\lambda]$ can be further truncated when $\lambda$ takes an integer value, but the truncation depends on whether the integer number is even or odd.
For $\lambda = 2 n +1$ and $\lambda = 2n$  $(n=1,2,\ldots)$, $hs^\text{e}[\lambda]$ can be reduced to $so(2 n+1)$ and $sp(2n)$, respectively \cite{Feigin_1988}.
In a similar way, we can show that there are two types of truncation with integer $\lambda$ for each restricted matrix extension, and hence we deal with four types of higher spin gravity with truncated spins in total.

For the truncated cases, we consider the Chern-Simons gravities with $\mathfrak{g}=so(M (2 n+1)), sp(2 Mn), sp(2 m(2 n+1)), so(4 m n)$. 
We further decompose the gauge algebras like \eqref{sldec} but with subalgebra $so(M)$ or $sp(2m)$, see table \ref{fig:hsa}. 
\begin{table}
	\centering
	\begin{align*}
	\begin{array}{ll}
	\text{Gauge algebra} ~ \mathfrak{g}  & \text{Decomposition} \\ \hline \hline
	so(M (2 n+1)) & so(M) \oplus so(2n+1) \oplus \cdots  \\ \hline 
	sp(2 M n) & so(M) \oplus sp(2n) \oplus \cdots \\ \hline 
	sp(2 m (2 n+1)) & sp(2m) \oplus so(2n+1) \oplus \cdots  \\ \hline 
	so(4 m n) & sp(2m) \oplus sp(2n) \oplus \cdots \\ \hline 
	osp(M (2 n+1)| 2 M n) & so(M) \oplus so(M) \oplus osp(2n+1| 2n) \oplus \cdots  \\ \hline 
	osp(M (2 n -1)| 2M n) & so(M) \oplus so(M) \oplus osp(2n -1 | 2n) \oplus \cdots \\ \hline 
	osp(4 m n | 2m (2 n +1)) & sp(2m) \oplus sp(2m) \oplus osp(2n+1| 2n) \oplus \cdots  \\ \hline 
	osp(4 m n | 2m (2 n -1)) & sp(2m) \oplus sp(2m) \oplus osp(2n-1|2n) \oplus \cdots \\ \hline
	\end{array} 
	\end{align*}
	\caption{8 types of restricted matrix extension for Chern-Simons gauge algebra. The gravitational (or supergravitational)  sector is identified with the principally embedded $sl(2)$ in $so(2n+1)$ or $sp(2n)$ (or $osp(1|2)$ in $osp(2 n \pm 1 |2n)$).}
	\label{fig:hsa}
\end{table}
With specified gravitational sectors, the asymptotic symmetries of the corresponding higher spin gravities are identified as the W-algebras obtained by the Hamiltonian reductions of $\mathfrak{g}$.
We may call the corresponding W-algebras as type $\mathfrak{g}$.
We compute the central charges $c$ and the levels $\ell$ of $so(M)$ or $sp(2m)$ currents from the Hamiltonian reductions.
The W-algebras of type $sp(2 M n),  so(4 m n)$ with $n=1$ include only generators of spin up to two, and we obtain the OPEs among the generators by requiring the associativity.  
We further conjecture that the rectangular W-algebras with $so(M)$ or $sp(2m)$ symmetry can be realized as the symmetry algebras of the cosets in table \ref{fig:proposal}.
\begin{table}
	\centering
	\begin{align*}
	\begin{array}{ll}
	\text{Type of W-algebra} & \text{Dual coset model} \\ \hline \hline
	\displaystyle so(M(2n+1)) & \displaystyle \frac{so(N+M)_k}{so(N)_k}  \\ \hline
	\displaystyle sp(2Mn) & \displaystyle \frac{osp(M|2N)_k}{sp(2N)_k} \\ \hline
	\displaystyle sp(2m(2n+1)) & \displaystyle \frac{osp(N|2m)_k}{so(N)_k} \\ \hline
	\displaystyle so(4mn) & \displaystyle \frac{sp(2N+2m)_k}{sp(2N)_k} \\ \hline
	\displaystyle osp(M (2 n+1)| 2 M n) & \displaystyle \frac{so(N+M)_k\oplus so(NM)_1}{so(N)_{k+M}} \\ \hline
	\displaystyle osp(M (2 n -1)| 2M n) & \displaystyle \frac{osp(M|2N)_k \oplus sp(2NM)_{-1/2}}{sp(2N)_{k-M/2}} \\ \hline
	\displaystyle osp(4 m n | 2m (2 n +1)) & \displaystyle \frac{osp(N|2m)_k \oplus sp(2 N m)_{-1/2}}{so(N)_{k-2m}}\\ \hline
	\displaystyle osp(4 m n | 2m (2 n -1)) & \displaystyle \frac{sp(2N+2m)_k \oplus so(4Nm)_1}{sp(2N)_{k+m}} \\ \hline
	\end{array} 
	\end{align*}
	\caption{Our proposals on the coset models whose symmetries are realized by the rectangular W-algebras. These W-algebras are obtained by the Hamiltonian reductions of the gauge (super)algebras and with the $sl(2)$ (or $osp(1|2)$) embeddings listed in table \ref{fig:hsa}.}
	\label{fig:proposal}
\end{table}
We provide the maps of parameters by comparing the central charges $c$ and the levels $\ell$ of the $so(M)$ or $sp(2m)$ currents. Moreover, we construct symmetry generators up to spin two and reproduce the OPEs obtained from the associativity.

The analysis can be generalized by introducing $\mathcal{N}=1$ supersymmetry.
As in the bosonic cases, we can construct the restricted matrix extensions of higher spin supergravity with $so(M)$ or $sp(2m)$ symmetry.  See \cite{Eberhardt:2018plx} for the holography with the higher spin supergravities.
Without matrix extensions, the holography with $\mathcal{N}=1$ truncated higher spin supergravity was proposed in \cite{Creutzig:2012ar}
 and the quantum asymptotic symmetry was examined in \cite{Candu:2013uya}.
For the cases with truncated spins, we use the Chern-Simons supergravities based on superalgebras $\mathfrak{g}=osp(M (2 n+1)|2Mn), osp(M(2n-1)|2 M n), osp(4 m n | 2 m (2 n +1)), osp(4 m n | 2 m (2 n-1))$ and decompose like \eqref{sldec} but with subalgebra $so(M) \oplus so(M)$ or $sp(2m) \oplus sp(2m)$ as in table \ref{fig:hsa}.
The asymptotic symmetries of the higher spin supergravities are the W-algebras obtained as the Hamiltonian reductions  as in the bosonic cases, and the corresponding W-algebras are again called as type $\mathfrak{g}$.
The central charges $c$ and the levels $\ell_1,\ell_2$ of $so(M)$ or $sp(2m)$ currents are computed, and
the OPEs among generators are obtained for the simple cases with currents of spin up to two by requiring associativity of the OPEs.
We also conjecture that the cosets in table \ref{fig:proposal} realize the $\mathcal{N}=1$ rectangular W-algebras as the symmetry algebras. We give the maps of parameters by comparing the central charges $c$ and the levels $\ell_1,\ell_2$ of the affine symmetries and explicitly construct low spin generators of the symmetry algebras of the cosets.

In the companion paper \cite{Creutzig:2019qos}, two of the current authors extend the previous analysis of \cite{Creutzig:2018pts} on the rectangular W-algebras with $su(M)$ symmetry in several ways. Firstly, we examine the OPEs among low spin generators with removing the restriction of  $n = 2$. Moreover, we study the degenerate representations of the rectangular W-algebra with $M = n =2$ and examine the relations to states of the coset \eqref{coset} and conical defect geometry of Chern-Simons gravity in \cite{Castro:2011iw} (see also \cite{Gaberdiel:2012ku,Perlmutter:2012ds,Hikida:2012eu}). It should be possible to apply these analyses to the current examples with restricted matrix extensions. In \cite{Creutzig:2019qos}, we also extend the previous analysis of \cite{Creutzig:2018pts} by introducing the $\mathcal{N}=2$ supersymmetry. 
It is important to consider the cases with more extended supersymmetry as in \cite{Creutzig:2014ula,Candu:2014yva,Hikida:2015nfa}.

This paper is organized as follows.
In the next section, we introduce the higher spin (super)algebras, which are used to construct higher spin (super)gravities. With the decompositions as listed in table \ref{fig:hsa}, we identify $sl(2)$ subalgebras as the gravitational sectors. 
In section \ref{sec:bosonic}, we examine the properties of the rectangular W-algebras with finite $c$ for the bosonic cases. We explain the spin contents of the W-algebras and write down the central charges $c$ and the levels $\ell$ of the $so(M)$ or $sp(2m)$ currents.
We then study the OPEs among generators for the simple examples with generators of spin up to two.
We further conjecture that the rectangular W-algebras can be realized as the symmetry algebras of the cosets listed in table \ref{fig:proposal} and provide some supports.
In section \ref{sec:super}, we extend the analysis by introducing $\mathcal{N}=1$ supersymmetry.
In appendix \ref{sec:candk}, we provide the derivations of the central charges and the levels of  $so(M)$ or $sp(2m)$ currents for the rectangular W-algebras in the both cases with and without  supersymmetry.
In appendix \ref{sec:alternative}, we propose alternative coset descriptions of the rectangular W-(super)algebras by swapping the dual coset models in table \ref{fig:proposal} and provide several computations that support our proposal.

\section{Higher spin (super)algebras}

Three dimensional higher spin (super)gravity can be described by Chern-Simons gauge fields $A, \bar A$.
Here $A$ and $\bar A$ take values in a higher rank gauge (super)algebra $\mathfrak{g}$.
For $\mathfrak{g}=sl(2)$, the Chern-Simons gauge theory describes the pure gravity on AdS$_3$ \cite{Achucarro:1987vz,Witten:1988hc}. For generic $\mathfrak{g}$, we should specify the gravitational $sl(2)$ subsector.
In this section, we introduce the restricted matrix extensions of $hs[\lambda]$ and its $\mathcal{N}=1$ supersymmetric extensions introduced in \cite{Prokushkin:1998bq}.
We then argue that the (super)algebras can be truncated when $\lambda$ takes an integer number just as  the algebra $hs_M[\lambda]$ can be reduced to $sl(Mn)$ at $\lambda=n$ $(n=2,3,\ldots)$  \cite{Creutzig:2018pts}.
Using the decomposition similar to \eqref{sldec},  we identify an $sl(2)$ subalgebra as the gravitational sector, see table \ref{fig:hsa}.
We examine the bosonic W-algebras in the next subsection and then generalize the analysis to the $\mathcal{N}=1$ W-algebras in subsection \ref{sec:hssa}.

\subsection{Restricted higher spin algebras}
\label{sec:hsa}

Without any matrix extensions and restrictions, the gauge algebra for the bosonic truncation of Prokushkin-Vasiliev theory of  \cite{Prokushkin:1998bq} is given by $hs[\lambda]$. 
In order to introduce the higher spin algebra, we define $B[\lambda]$ by the universal enveloping algebra of $sl(2)$ with removing an ideal as
\begin{align}
B[\lambda] = \frac{U(sl(2))}{\langle C_2 - \frac14 (\lambda^2 -1) \mathbbm{1} \rangle} \, .
\end{align}
Here $C_2$ is the Casimir of $sl(2)$.
The generators of $B[\lambda]$ may be expressed as $V^1_0 = \mathbbm{1}$ and $V^s_m$ with $s=2,3,\ldots$ and $m = - s+ 1 , -s  + 2 , \ldots , s-1$.
The higher spin algebra $hs[\lambda]$ is then given by removing $V^1_0 = \mathbbm{1}$ from $B[\lambda]$ as
\begin{align}
B[\lambda] = hs[\lambda] \oplus \mathbbm{1} \, .
\end{align}
The higher spin algebra satisfies a relation $hs[\lambda] = hs[- \lambda]$.
At $\lambda = n$, the algebra $hs[\lambda]$ can be truncated to $sl(n)$ by diving an ideal.
Multiplying the $M \times M$ matrix algebra, we can define $hs_M[\lambda]$ as
\begin{align}
gl(M) \otimes B[\lambda] = hs_M [\lambda] \oplus \mathbbm{1}_M \otimes \mathbbm{1} \, .
\end{align}
An $M \times M$ matrix is expressed by $A_M$ with an index parameterizing its size, and, in particular, 
the $M \times M$ identity matrix is denoted as $\mathbbm{1}_M$.
With the terminology, the generators of $gl(M) \otimes B[\lambda]$ is expressed by 
$A_M \otimes V^s_m$. 
The algebra $hs_M[\lambda]$ can be decomposed as
\begin{align}
 hs_M[\lambda] \simeq sl(M) \otimes \mathbbm{1} \oplus \mathbbm{1}_M \otimes hs[\lambda] \oplus sl(M) \otimes hs[\lambda] \, .
\end{align}
From this decomposition, we can see that the algebra can be truncated to $sl (M n)$ as
\begin{align}
sl(Mn) \simeq sl(M) \otimes \mathbbm{1}_n \oplus \mathbbm{1}_M \otimes sl(n) \oplus sl(M) \otimes sl(n) 
\end{align}
at $\lambda = n$.
We use the principally embedded $sl(2)$ in $\mathbbm{1}_M \otimes hs[\lambda] $ or $ \mathbbm{1}_M \otimes sl(n)$ as the gravitational sector.

We can further generalize the gauge algebras by putting restrictions on the extra matrix degrees of freedom,
and it is known that there are two ways to do so \cite{Prokushkin:1998bq}, see also \cite{Eberhardt:2018plx}.
For one type, we decompose the $M \times M$ matrix $A_M$ as
\begin{align}
A_M^a: A^\text{tr}_M = - A_M  \, , \quad A_M^s:  A^\text{tr}_M =  A_M\, , \label{Aas}
\end{align}
where $A^\text{tr}_M$ represents the matrix transpose of $A_M$. 
We further decompose the symmetric part as $A_M^s = \mathbbm{1}_M \oplus A_M^{s'}$ by decoupling the trace part. 
A restricted algebra is then given by
\begin{align}
A_M^a \otimes \mathbbm{1} \oplus \mathbbm{1}_M \otimes hs^e [\lambda] \oplus  A_M^{s'} \otimes hs^\text{e} [\lambda]  \oplus A_M^a \otimes hs^\text{o} [\lambda] \ , \label{exthse1}
\end{align} 
where $hs^\text{e} [\lambda]$  and $hs^\text{o} [\lambda]$ are even and odd spin parts of $hs[\lambda]$, respectively.
From this expression, we can see that the algebra includes $so(M)$ and $hs^\text{e}[\lambda]$ as subalgebras.
For $M = 2m$ with $m=1,2,\ldots$, there is another type of restriction.
In this case, we consider a different type of decomposition as
\begin{align}
A_{2m}^{\Omega , a}: \Omega_{2m}^{-1} A^\text{tr}_{2m} \Omega_{2m} = - A_{2m}  \, , \quad A_{2m}^{\Omega ,s }: \Omega_{2m}^{-1} A^\text{tr}_{2m} \Omega_{2m} = A_{2m} \, , \quad \Omega^\text{tr}_{2m} = - \Omega_{2m} \, . \label{AOmegaas}
\end{align}
Here $\Omega_{2m}$ is a symplectic form expressed by a $2 m \times 2m$ matrix.
We can decouple the trace part as $A^{\Omega , s}_{2m} = \mathbbm{1}_{2m} \oplus A^{\Omega , s '}_{2m}$.
With these decompositions, we can introduce anther restricted algebra as
\begin{align}
A_{2m}^{\Omega ,a} \otimes \mathbbm{1} \oplus  \mathbbm{1}_{2m} \otimes hs^\text{e} [\lambda] \oplus  A_{2m}^{\Omega ,s '} \otimes hs^\text{e} [\lambda]  \oplus A_{2m}^{\Omega ,a } \otimes hs^\text{o} [\lambda] \ . \label{exthse2}
\end{align} 
Since $A_{2m}^{\Omega ,a}$ generates $sp(2m)$, the algebra includes $sp(2m)$ along with $hs^\text{e}[\lambda]$ as subalgebras.
For the both cases in \eqref{exthse1} and \eqref{exthse2}, we use the principally embedded $sl(2)$ in $\mathbbm{1}_{M} \otimes hs^\text{e} [\lambda]$ (and with $M=2m$) as the gravitational sector.

In order to analyze  full quantum corrections, it is convenient to work with an integer $\lambda$, where the algebras can be truncated by dividing an ideal.
At $\lambda = 2 n+1$ $(n=1,2,\ldots)$, $hs^\text{e} [\lambda]$ can be truncated to $so(2 n+1)$, which may be
 generated by $B^{a}_{2n+1}: B_{2 n+1}^\text{tr} = - B_{2n +1}$.  Similarly, $hs^o [\lambda]$ may be expressed by $B^{s'}_{2n+1}$ at the value of $\lambda$. This may be  understood from
\begin{align}
hs[\lambda] \simeq hs^\text{e} [\lambda] \oplus hs^\text{o} [\lambda] \, ,
\end{align}
where the generators of $hs[\lambda]$ with $\lambda = 2n +1$ could be given by traceless $(2n +1) \times (2n +1)$ matrices.
Thus, at $\lambda = 2 n+1$, we have the truncated algebras
\begin{align} \label{talgebra0}
&A_M^a \otimes \mathbbm{1}_n \oplus 
\mathbbm{1}_M \otimes B_{2 n +1}^a \oplus 
A_M^{s'} \otimes B_{2 n +1}^a  \oplus A_M^a \otimes B_{2 n+1}^{s'} \ , \\
&A_{2m}^{\Omega ,a} \otimes \mathbbm{1}_n \oplus \mathbbm{1}_{2m} \otimes B_{2 n +1}^a \oplus A_{2m}^{\Omega ,s'} \otimes B_{2 n +1}^a  \oplus A_{2m}^{\Omega ,a} \otimes B_{2 n+1}^{s'}  \ .
\label{talgebra1}
\end{align} 
The first algebra includes $so(M)$ and $so(2 n+1)$ as subalgebras, while
the second one includes  $sp(2 m)$ and $so(2 n+1)$ as subalgebras.
The gravitational $sl(2)$ is identified with the one principally embedded in the $so(2n+1)$.
Similarly, at $\lambda = 2n$,  $hs^\text{e} [\lambda]$ can be truncated to $sp(2n)$, which implies that $hs^\text{e} [\lambda]$ and $hs^\text{o} [\lambda]$ may be generated $B_{2 n }^{\Omega ,a}$ and $B_{2n}^{\Omega ,s '}$, respectively. 
The truncated algebras at $\lambda = 2n$ are
\begin{align}
\label{talgebra2}
&A_M^a \otimes \mathbbm{1}_{2n} \oplus \mathbbm{1}_M \otimes B_{2 n}^{\Omega ,a} \oplus A_M^{s '} \otimes B_{2 n}^{\Omega ,a}  \oplus A_M^a \otimes B_{2 n}^{\Omega ,s'}  \ , \\
&A_{2m}^{\Omega,a} \otimes \mathbbm{1}_{2 n} \oplus \mathbbm{1}_{2m} \otimes B^{\Omega,a}_{2 n} \oplus A_{2m}^{\Omega ,s'} \otimes B_{2 n}^{\Omega ,a}  \oplus A_{2m}^{\Omega ,a} \otimes B_{2 n}^{\Omega ,s'}  \ .
\label{talgebra3}
\end{align} 
The first algebra includes $so (M)$ and $sp (2 n)$ as subalgebras, and the second one includes $sp(2m)$ and $sp(2n)$ as subalgebras. 
The gravitational $sl(2)$ is identified with the one principally embedded in the $sp(2n)$.

The expressions of restricted algebras in \eqref{talgebra0},  \eqref{talgebra1}, \eqref{talgebra2} and \eqref{talgebra3} look complicated.
However, as seen below, they can be regarded as decompositions of Lie algebras $so(ML)$ and $sp(2 m L)$ with $L=2n$ or $L=2n +1$, see table \ref{fig:hsa}.
We first consider $so(ML)$ and express the generators by $T_{ML}$, which  satisfy
\begin{align}
\Omega_{ML}^{-1}T_{ML}^\text{tr} \Omega_{ML} = - T_{ML} \, , \quad
\Omega_{ML}^\text{tr} = \Omega_{ML} \, .
\label{socond0}
\end{align}
We then choose to express $T_{ML}$ by tensor products $A_M \otimes B_L$.
With this expression, the condition \eqref{socond0} can be realized by
\begin{align}
\Omega^{-1}_M A_M^\text{tr} \Omega_M  = \epsilon A_M\, , \quad
\Omega^{-1}_L B_L^\text{tr} \Omega_L  = - \epsilon B_L
\end{align}
and
\begin{align}
\Omega_M^\text{tr} = \eta  \Omega_M \, , \quad
\Omega_L^\text{tr} = \eta \Omega_L
\end{align}
with $\epsilon = \pm 1$ and $\eta = \pm 1$.
Here $\eta = - 1$ is possible only for $M = 2m$ and $L=2n$.
We can see that the cases with $\eta = 1$ and $\eta = -1$ correspond to \eqref{talgebra0} and \eqref{talgebra3}, respectively.
We then examine $sp(ML)$ with generators $T_{ML}$ satisfying
\begin{align}
\Omega_{ML}^{-1}T_{ML}^\text{tr} \Omega_{ML} = - T_{ML} \, , \quad
\Omega_{ML}^\text{tr} = - \Omega_{ML} \, .
\label{spcond0}
\end{align}
Here we have assumed that $ML$ is even. 
As above, $T_{ML}$ are expressed by tensor products $A_M \otimes B_L$ and
the condition \eqref{spcond0} can be realized by
\begin{align}
\Omega^{-1}_M A_M^\text{tr} \Omega_M  = \epsilon A_M\, , \quad
\Omega^{-1}_L B_L^\text{tr} \Omega_L  = - \epsilon B_L
\end{align}
and
\begin{align}
\Omega_M^\text{tr} = \eta  \Omega_M \, , \quad
\Omega_L^\text{tr} = - \eta \Omega_L 
\end{align}
with $\epsilon = \pm 1$ and $\eta = \pm 1$.
Here $\eta = - 1$ is possible if $M = 2m$, and the case corresponds to \eqref{talgebra1}. Similarly, $\eta = + 1$ is possible if $L = 2n$, and the case corresponds to \eqref{talgebra2}.

\subsection{Restricted higher spin superalgebras}
\label{sec:hssa}

The gauge algebra for the $\mathcal{N}=2$ Prokushkin-Vasiliev theory of \cite{Prokushkin:1998bq} is given by $shs[\lambda]$ without matrix extensions and restrictions.
As in the case of $hs[\lambda]$, we introduce $sB[\lambda]$ by the universal enveloping algebra of $osp(1|2)$ divided by an ideal as 
\begin{align}
sB[\lambda] = \frac{U(osp(1|2))}{\langle  C_2 - \frac14 \lambda (\lambda -1) \mathbbm{1} \rangle} \, .
\label{sB}
\end{align}
Here $C_2$ is the Casimir operator of $osp(1|2)$. Decoupling the identity element $\mathbbm{1}$, we define a higher spin superalgebra $shs[\lambda]$ as \cite{Bergshoeff:1991dz}
\begin{align}
sB[\lambda] = shs[\lambda] \oplus \mathbbm{1} \, . 
\end{align}
The higher spin superalgebra satisfies a relation $shs[\lambda] = shs[1 -\lambda]$.
At $\lambda = - n $, the superalgebra can be truncated to $sl(n +1 | n )$.
Multiplying the $M \times M$ matrix algebra, we define $shs_M[\lambda]$ by (see \cite{Gaberdiel:2013vva,Creutzig:2013tja})
\begin{align}
gl(M) \otimes sB[\lambda] = shs_M[\lambda] \oplus \mathbbm{1}_M \otimes \mathbbm{1} \, .
\end{align}
The superalgebra can be decomposed as
\begin{align}
 shs_M[\lambda] \simeq sl(M) \otimes \mathbbm{1} \oplus \mathbbm{1}_M  \otimes  shs[\lambda] \oplus   sl(M)  \otimes shs[\lambda] \, ,
\end{align}
which can be truncated to
\begin{align}
sl(M (n+1)| Mn) \simeq   sl(M) \otimes \mathbbm{1} \oplus \mathbbm{1}_M \otimes sl(n+1|n) \oplus sl(M) \otimes sl(n+1|n) 
\end{align}
at $\lambda = - n$.
We can describe $\mathcal{N}=1$  supergravity on AdS$_3$ by $osp(1|2)$ Chern-Simons gauge theory \cite{Achucarro:1987vz}, and it is important to identify the supergravity sector in order to define a  higher spin supergravity. In the current case, we use the principal embedding of $osp(1|2)$ in $\mathbbm{1}_M \otimes shs[\lambda] $ or $\mathbbm{1}_M \otimes sl(n+1|n) $

The superalgebra $shs[\lambda]$ has a symmetry under a $\mathbb{Z}_2$ action $\sigma$ as closely explained in \cite{Candu:2013uya}, and the $\mathbb{Z}_2$ truncation of $shs[\lambda]$ is denoted as $shs^\sigma [\lambda]$.
There are the sectors of integer and half-integer spins, and the truncated algebra includes only even spin generators for the integer spin sector.
As explained in \cite{Candu:2013uya}, $shs^\sigma [\lambda]$ can be truncated to $osp(2n+1|2n)$ at $\lambda = - 2n$ and $osp(2n-1|2n)$ at $\lambda = - 2n + 1$ $(n=1,2,\ldots)$.
Just as in the bosonic case, there are two types of truncation for the matrix extension as
\begin{align}
\begin{aligned}
&A_M^a \otimes \mathbbm{1} \oplus  \mathbbm{1}_M \otimes shs^\sigma [\lambda] \oplus  A_M^{s'} \otimes shs^\sigma [\lambda]  \oplus A_M^a \otimes shs^{\bar \sigma} [\lambda] \ , \\
&A_{2m}^{\Omega ,a} \otimes \mathbbm{1} \oplus  \mathbbm{1}_{2m} \otimes shs^\sigma [\lambda]  \oplus  A_{2m}^{\Omega ,s '} \otimes shs^\sigma [\lambda]  \oplus A_{2m}^{\Omega ,a} \otimes shs^{\bar \sigma } [\lambda] \, .
\end{aligned}
\label{tsalgebra}
\end{align} 
Here we have used the notations introduced in \eqref{Aas} and \eqref{AOmegaas}.
Moreover, we decompose $shs[\lambda]$ as
\begin{align}
shs[\lambda] \simeq shs^\sigma [\lambda] \oplus shs^{\bar \sigma} [\lambda] \, ,
\end{align}
where $shs^{\bar \sigma} [\lambda] $ includes only odd spin generators for the integer spin sector.
The first and second algebras in \eqref{tsalgebra} include $so(M)$ and $sp(2m)$ subalgebras, respectively.
As the supergravity sector, we use the $osp(1|2)$ principally embedded in $\mathbbm{1}_M \otimes shs^\sigma [\lambda]$ or with $M=2m$.

We would like to discuss the truncation of extended algebras at $\lambda = - 2n$ or $\lambda = - 2n +1$, where $shs^{\sigma}[\lambda]$ reduces  $osp(2n+1|2n)$ or $osp(2n-1|2n)$.
For this, we start by reviewing some properties of supermatrices and  Lie superalgebra $osp( m | 2n)$, see, e.g., \cite{Frappat:1996pb}. We express the generators of $osp( m | 2n)$ by even supermatrices of the form
\begin{align}
M _{m|2n}= \begin{pmatrix}
A_m & B_{m,2n} \\
C_{2n,m} & D_{2n}
\end{pmatrix} \, ,
\end{align}
where $A_m , B_{m,2n} ,C_{2n,m} ,D_{2n} $ are $m \times m$, $m \times 2n$, $2n \times m$, $2n \times 2n$ matrices, respectively.
We define the supertrace and the supertranspose as
\begin{align}
\text{str} M _{m|2n}= \text{tr} A_m - \text{tr} D_{2 n} \, , \quad
M^\text{str} _{m|2n} = \begin{pmatrix}
A_m^\text{tr} & C_{2n,m}^\text{tr} \\
- B_{m,2n}^\text{tr} & D_{2n}^\text{tr} 
\end{pmatrix} \, . \label{str}
\end{align}
The condition for the generators of $osp( m | 2n)$ is then given by
\begin{align}
\begin{pmatrix}
\Omega_{m}^{-1} &0 \\
0  & \Omega_{2n} ^{-1}
\end{pmatrix} M^\text{str}_{m|2n} \begin{pmatrix}
\Omega_m &0 \\
0  & \Omega_{2n} 
\end{pmatrix}= - M _{m|2n}  \, , 
\quad \Omega_m^\text{tr} = \Omega_m \,  , \quad  \Omega_{2n} ^\text{tr} = - \Omega_{2 n} \, ,
\label{ospcond1}
\end{align}
which can be rewritten as
\begin{align}
\Omega_m^{-1} A^\text{tr}_m \Omega_m = - A_m \, , \quad \Omega_{2n}^{-1} D^\text{tr}_{2n} \Omega_{2n} = - D_{2n} \, , \quad \Omega_{2n}^{-1} B^\text{tr}_{m,2n} \Omega_{m} = C_{2n,m} \, .
\label{ospcond2}
\end{align}
From this expression, we can see that $A_m$ and $D_{2n }$ generates $so(m)$ and $sp(2n)$, respectively.

We express a supermatrix satisfying the condition of $osp(m|2n)$ by $B_{m|2n}^\sigma$.
We then schematically decompose a supermatrix $B_{m|2n}$  as
\begin{align}
B_{m|2n} = \mathbbm{1}_{m|2n} \oplus B_{m|2n}^\sigma \oplus B_{m|2n}^{\bar \sigma} \, ,
\end{align}
where $ \mathbbm{1}_{m|2n} $ actually means
\begin{align}
\mathbbm{1}_{m|2n} = 
\begin{pmatrix}
\mathbbm{1}_m & 0 \\
0 & 0 
\end{pmatrix}
\oplus
\begin{pmatrix}
0& 0 \\
0 & \mathbbm{1}_{2n}
\end{pmatrix} \, . \label{dec1m2n}
\end{align}
With this terminology, the expressions in \eqref{tsalgebra} can be truncated to
\begin{align}
&A_M^a \otimes \mathbbm{1}_{2n+1|2n} \oplus  \mathbbm{1}_M \otimes B_{2n+1|2n}^\sigma \oplus  A_M^{s'} \otimes B_{2n+1|2n}^\sigma \oplus A_M^a \otimes B_{2n+1|2n}^{\bar \sigma}  \ , \label{tsalgebra0}\\
&A_{2m}^{\Omega ,a} \otimes \mathbbm{1}_{2n+1|2n}   \oplus  \mathbbm{1}_{2m} \otimes  B_{2n+1|2n}^\sigma\oplus  A_{2m}^{\Omega ,s '} \otimes  B_{2n+1|2n}^\sigma \oplus A_{2m}^{\Omega ,a} \otimes B_{2n+1|2n}^{\bar \sigma} \label{tsalgebra1}
\end{align} 
at $\lambda = -2 n$. 
Noticing \eqref{dec1m2n}, we can see that
the expressions \eqref{tsalgebra0} and \eqref{tsalgebra1} include two sets of $so(M)$ and $sp(2m)$ as subalgebras.
Moreover, we can show as in appendix \ref{sec:candks} that they are identified as specific decompositions of $osp(M(2n+1)|2 Mn)$ and $osp(4mn|2m(2n+1))$, respectively.
Similarly, the expressions in \eqref{tsalgebra} can be truncated to
\begin{align}
&A_M^a \otimes \mathbbm{1}_{2n-1|2n } \oplus  \mathbbm{1}_M \otimes  B_{2n-1|2n}^\sigma \oplus  A_M^{s'} \otimes  B_{2n-1|2n}^\sigma \oplus A_M^a \otimes  B_{2n-1|2n}^{\bar \sigma}  \ ,  \label{tsalgebra2} \\
&A_{2m}^{\Omega ,a} \otimes  \mathbbm{1}_{2n-1|2n} \oplus  \mathbbm{1}_{2m} \otimes B_{2n-1|2n}^\sigma \oplus  A_{2m}^{\Omega ,s '} \otimes B_{2n-1|2n}^\sigma \oplus A_{2m}^{\Omega ,a} \otimes B_{2n-1|2n}^{\bar \sigma}  \label{tsalgebra3}
\end{align} 
at $\lambda = - 2n +1$. The expressions \eqref{tsalgebra2} and \eqref{tsalgebra3} include  two sets of $so(M)$ and $sp(2m)$ as subalgebras and are identified as specific decompositions of $osp(M(2n-1)|2 M n)$ and $osp(4mn|2m(2n-1))$, respectively, see again appendix \ref{sec:candks}.

\section{Rectangular W-algebras}
\label{sec:bosonic}

In subsection \ref{sec:hsa}, we constructed four types of restricted matrix extensions for the gauge algebra and the gravitational $sl(2)$ sectors are specified, see table \ref{fig:hsa}.
In order to construct a Chern-Simons gravity theory, we need to assign an asymptotic AdS condition.
We choose the asymptotic AdS condition such that
\begin{align}
\left. (A - A_\text{AdS} ) \right |_{\rho \to \infty} = \mathcal{O} (1) \, . 
\label{AdScond}
\end{align}
Here $\rho$ is the radial coordinate of AdS$_3$ and the AdS boundary is located at $\rho \to \infty$. Moreover, $A_\text{AdS}$ denotes the configuration of gauge fields describing the AdS background.
With this setup, the asymptotic symmetry is given by the Hamiltonian reduction of the gauge algebra with the $sl(2)$ embedding \cite{Henneaux:2010xg,Campoleoni:2010zq}.
In this way, we now have four types of rectangular W-algebras, and we may call them as types $so(M(2n+1)),sp(2 Mn), sp(2m (2 n +1)), so(4 m n)$.
In the next subsection, we explain the basic properties of the W-algebras including spin contents, central charges, and the levels of affine $so(M)$ or $sp(2m)$ algebra. In subsection \ref{sec:OPEs}, we compute OPEs  for simple examples involving only generators up to spin two.
In subsection \ref{sec:map}, we propose the dual cosets whose symmetries are the same W-algebras  as listed in table \ref{fig:proposal} and provide the maps of parameters. In subsection \ref{sec:cosetalgabra}, we explicitly construct the generators of the symmetry algebras up to spin two and compare the OPEs obtained in subsection \ref{sec:OPEs}.

\subsection{Basic properties of the algebras}
\label{sec:basic}

In this subsection, we examine the basic properties of the rectangular W-algebras obtained by  quantum Hamiltonian reductions of the gauge algebras listed in table \ref{fig:hsa}.
We first obtain the spin contents of the W-algebras and then write down the central charges $c$ of the algebras and the levels $\ell$ of the $so(M)$ or $sp(2m)$ currents. 
We explain the derivations of $c$ and $\ell$ in appendix \ref{sec:candkb}.

\subsubsection*{\underline{Type $so(M(2n+1))$}}

We set the gauge algebra as $\mathfrak{g}=so(M(2n+1))$ and decompose it as (see \eqref{talgebra0})
\begin{align}
\begin{aligned}
so(M (2 n+1))\simeq \, &  so(M) \otimes \mathbbm{1}_{2n+1} \oplus   \mathbbm{1}_{M} \otimes so (2 n +1)   \\
&\oplus (\text{sym})_M \otimes so(2 n+1) \oplus so(M) \otimes (\text{sym})_{2 n+1} \, .
\end{aligned} \label{decsoI}
\end{align}
Here $(\text{sym})_L$ denotes the symmetric representation of $so(L)$, which are generated by the matrices $A^{s'}_L$. We use  the principally embedded $sl(2)$ in  $\mathbbm{1}_{M} \otimes so (2 n +1) $ as the gravitational $sl(2)$. With the identification of gravitational sector, spin $s$ for the embedded $sl(2)$ can be directly related to the space-time spin. 
We denote the elements of $sl(2)$ by the triplet $\{\hat x,\hat e,\hat f\}$ satisfying
\begin{align}
[\hat x ,\hat e] = \hat e \, , \quad [\hat x, \hat f] = - \hat  f \, , \quad  [\hat e ,\hat  f] = \hat x \, .
\label{sl2comm}
\end{align}
With the $sl(2)$ action, we can decompose $so(2n +1)$ and $(\text{sym)}_{2 n +1}$ in \eqref{decsoI} as
\begin{align}
so(2 n +1) = \bigoplus_{s=1}^{n} \rho_{2s -1}\, , \quad (\text{sym})_{2 n + 1} = \bigoplus_{s=1}^{n} \rho_{2s}  \, , \label{sosymdec}
\end{align}
where $\rho_s$ denotes the spin $s$ representation of $sl(2)$. 
Using the dimensions of $so(M)$ and $\mathbbm{1}_M \oplus (\text{sym})_M$ as
\begin{align}
C_M^a = \frac{M (M-1)}{2} \, , \quad C_M^s = \frac{M (M+1)}{2} \,  ,
\label{CM}
\end{align}
the gauge algebra $so(M (2 n+1))$ can be decomposed as
\begin{align}
\begin{aligned}
&so(M (2 n+1)) \simeq \frac{M (M-1)}{2} \left(\bigoplus_{s=0}^{n} \rho_{2s} \right) \oplus   \frac{M (M+1)}{2} \left(\bigoplus_{s=1}^{n} \rho_{2s -1} \right) \, .
\end{aligned}
\end{align}

We can see that the modes with positive eigenvalues of $\hat x$ diverge for $\rho \to \infty$, 
thus the asymptotic AdS condition \eqref{AdScond} requires that the positive modes vanish.
Moreover, we can set the other modes to be zero as well except for one element in each $\rho_s$.
Since the gauge fields $A=A_\mu d x^\mu$ have already one space-time index, the element has total space-time spin $s+1$ for the one from $\rho_s$.
Therefore, the algebra includes $M(M-1)/2$ odd spin currents with $s=1,3,\ldots, 2n+1$ and $M(M+1)/2$ even spin currents with $s=2,4,\ldots, 2n$ after the Hamiltonian reduction with the $sl(2)$ embedding. If we set $M=1$, then only even spin currents with $s=2,4,\ldots, 2n$ are left.
The spin content reproduces the one in \cite{Candu:2012ne} without matrix extension.

The  $M(M-1)/2$ spin one currents form  $so(M)$  affine   algebra, and the spin two current in the singlet of $so(M)$ is the energy-momentum tensor. In appendix \ref{sec:candkb}, the level of $so(M)$ currents is computed  as
\begin{align}
\ell = t (2 n+1) + ( M + 2 )2  n^2 +  (M-2) 2 n (n+1) \, , \label{lsoI}
\end{align}
and the central charge  is obtained as
\begin{align}
\begin{aligned}
c =& \frac{ tM (2 n+1) (M (2 n+1)-1)}{2 (t + M (2 n+1)-2)} - 2t M n (n+1) (2 n+1) \\ & +  \frac{M(M+1) }{2}  \left(6 n^2-8 n^4\right)-  \frac{M (M-1)}{2}2 n (n+1) (4 n (n+1)-1) \, .
\end{aligned} \label{csoI}
\end{align}
Here $t$ is the level of $so(M(2 n +1))$ affine algebra.
There could be several constructions of the same W-algebra, and the label $t$ is not an intrinsic parameter. 
We can obtain the relation between $c$ and $\ell$ as
\begin{align}
c = \frac{\ell M (-4 (\ell -1) n (n+1)+ M ( 2 n+ 1)-1)}{2 (\ell + M (2 n +1)-2)} 
\end{align}
by removing the parameter $t$.

\subsubsection*{\underline{Type $sp(2 M n )$}}

The gauge algebra is set as $\mathfrak{g}=sp(2 M n )$ and the decomposition is  (see \eqref{talgebra2})
\begin{align}
\begin{aligned}
sp(2 M n )\simeq \, &  so(M) \otimes \mathbbm{1}_{2n} \oplus   \mathbbm{1}_{M} \otimes sp (2 n)   \\
&\oplus (\text{sym})_M \otimes sp(2 n ) \oplus so(M) \otimes (\text{asym})_{2 n} \, .
\end{aligned}  \label{decsoII}
\end{align}
Here $(\text{asym})_{2 n}$ denotes the anti-symmetric representation of $sp(2n)$  generated by  matrices $A^{\Omega,s'}_{2n}$. As the gravitational sector, we use  the principally embedded $sl(2)$ in  $\mathbbm{1}_{M} \otimes sp (2 n) $. 
With the $sl(2)$, we decompose $sp(2 M n )$ as
\begin{align} 
\begin{aligned}
&sp(2 M n ) \simeq \frac{M (M-1)}{2} \left(\bigoplus_{s=0}^{n-1} \rho_{2s} \right) \oplus   \frac{M (M+1)}{2} \left(\bigoplus_{s=1}^{n} \rho_{2s -1} \right) \, .
\end{aligned}
\end{align}
Here we have used the decompositions of $sp(2n)$ and $(\text{asym})_{2n}$ in \eqref{decsoII} as
\begin{align}
sp(2n) = \bigoplus_{s=1}^{n} \rho_{2s -1} \, , \quad (\text{asym})_{2n } = \bigoplus_{s=1}^{n-1} \rho_{2s} \label{spasymdec}
\end{align}
as well as \eqref{CM}.
After the Hamiltonian reduction,  there are $M(M-1)/2$ odd spin currents with $s=1,3,\ldots,2n-1$ and $M(M+1)/2$ even spin currents with $s=2,4,\ldots, 2n$.
For $M=1$, the algebra includes only even spin currents with $s=2,4,\ldots, 2n$ as in \cite{Candu:2012ne}.

The  $M(M-1)/2$ spin one currents form $so(M)$ affine algebra, and the spin two current in the singlet of $so(M)$ is the energy-momentum tensor. In appendix \ref{sec:candkb}, the level of $so(M)$ currents is computed  as
\begin{align}
\ell = 2 t \cdot 2 n +  (M+2)2 n^2 + (M-2) 2 n (n-1)\, , \label{lsoII}
\end{align}
and the central charge  is obtained as
\begin{align}
\begin{aligned}
c =& \frac{t M n  (2 M n+1)}{t+M n+1} - t 2 M n \left(4 n^2-1\right)  \\
 &+\frac{ M (M+1)}{2} \left(6 n^2-8 n^4\right)- \frac{(M-1) M}{2} 2 n \left(4 n^3-8 n^2+3 n+1\right) \, .
\end{aligned}  \label{csoII}
\end{align}
Here the level of $sp(2 M n )$ affine algebra is denoted as $t$.
The central charge can be written in terms of $\ell $ as
\begin{align}
c =\frac{M \left(\ell^2+2 \ell M n-4 (\ell-2) (\ell-1) n^2\right)}{2 (\ell +2 M n)} \label{centersoII}
\end{align}
by removing $t$.

\subsubsection*{\underline{Type $sp(2 m (2 n +1) )$}}

We set $\mathfrak{g}=sp(2 m (2 n +1) )$ and decompose it as  (see \eqref{talgebra1})
\begin{align}
\begin{aligned}
sp(2 m (2 n +1) )\simeq \, &  sp(2 m) \otimes \mathbbm{1}_{2n+1} \oplus   \mathbbm{1}_{2 m} \otimes so (2 n + 1)   \\
&\oplus (\text{asym})_{2m} \otimes so(2 n +1 ) \oplus sp(2m) \otimes (\text{sym})_{2 n+1} \, .
\end{aligned}  \label{decspI}
\end{align}
We adopt the $sl(2)$ principally embedded in  $\mathbbm{1}_{2m} \otimes so (2 n +1) $.
The decomposition of $sp(2 m  (2 n +1) ) $ by the $sl(2)$ is
\begin{align} 
\begin{aligned}
&sp(2 m  (2 n +1) ) \simeq (2 m^2 + m)\left(\bigoplus_{s=0}^{n} \rho_{2s} \right) \oplus   (2 m^2 - m)\left(\bigoplus_{s=1}^{n} \rho_{2s -1} \right) \, ,
\end{aligned}
\end{align}
where we have used the dimensions of $sp(2m)$ and $\mathbbm{1}_{2m} \oplus (\text{asym})_{2m}$ as
\begin{align}
C_{2m}^{\Omega,a} = 2 m^2 + m \, , \quad C_{2m}^{\Omega,s} = 2 m^2 - m \label{C2m}
\end{align}
in addition to  \eqref{sosymdec}.
The algebra after the Hamiltonian reduction has $2 m^2 + m$ odd spin currents with $s=1,3,\ldots,2n+1$ and $2 m^2 - m$ even spin currents with $s=2,4,\ldots, 2n$. 
There appears no analogous algebra in \cite{Candu:2012ne} since the matrix extension is essential in this case.

The  $2 m^2 + m$ spin one currents form  $sp(2m)$ affine algebra, and the spin two current in the singlet of $sp(2m)$ is the energy-momentum tensor. As obtained in appendix \ref{sec:candkb}, the level of $sp(2m)$ currents is 
\begin{align}
\ell = t (2 n+1) + (m-1) 2 n^2 +  (m+1) 2 n (n + 1)\, ,  \label{lspI}
\end{align}
and the central charge  is 
\begin{align}
\begin{aligned}
&c = \frac{t m (2 n+1)  (2 m (2 n+1)+1)}{t + m (2 n+1)+1} -8 t m n (n+1) (2 n+1)  \\
& + ( 2 m^2 - m)  (6 n^2 - 8 n^4)  - ( 2 m^2 + m) 2 n (n + 1) ( 4 n ( n +1) -1) \, .
\end{aligned} \label{cspI}
\end{align}
Here $t$ is the level of $sp(2 m (2 n+1) )$ affine algebra.
Removing $t$, we have the relation between $c$ and $\ell$ as
\begin{align}
c = \frac{\ell m (-4 (2 \ell +1) n (n+1)+ 2 m (2 n+1)+1)}{\ell + m (2 n+1)+1} \, .
\end{align}

\subsubsection*{\underline{Type $so(4 m n)$}}

We set $\mathfrak{g}=so(4 m n)$ and use the decomposition  (see \eqref{talgebra3})
\begin{align}
\begin{aligned}
so(4 m n)\simeq \,  &  sp(2 m) \otimes \mathbbm{1}_{2n} \oplus   \mathbbm{1}_{2 m} \otimes sp (2 n)   \\
&\oplus (\text{asym})_{2m} \otimes sp(2 n) \oplus sp(2m) \otimes (\text{asym})_{2 n} \, .
\end{aligned}   \label{decspII}
\end{align}
The gravitational $sl(2)$ is identified with  the $sl(2)$ principally embedded in  $\mathbbm{1}_{2m} \otimes sp (2 n) $.  With the $sl(2)$, we decompose
\begin{align} 
\begin{aligned}
&so(4 m n ) \simeq (2 m^2 + m)\left(\bigoplus_{s=0}^{n-1} \rho_{2s} \right) \oplus   (2 m^2 - m)\left(\bigoplus_{s=1}^{n} \rho_{2s -1} \right) 
\end{aligned}
\end{align}
by using \eqref{spasymdec} and \eqref{C2m}.
After the Hamiltonian reduction, there are $2 m^2 + m$ odd spin currents with $s=1,3,\ldots,2n-1$ and $2 m^2 - m$ even spin currents with $s=2,4,\ldots, 2n$.

The  $2 m^2 + m$ spin one currents generate $sp(2m)$ affine algebra, and the spin two current in the singlet of $sp(2m)$ is the energy-momentum tensor. In appendix \ref{sec:candkb}, we obtain the level of $sp(2m)$ currents as
\begin{align}
\ell = t \cdot 2 n /2 +  (m-1)2  n^2 +  (m+1) 2 n (n - 1)\, ,  \label{lspII}
\end{align}
and the central charge  as
\begin{align}
\begin{aligned}
c =& \frac{t 2 m n  (4 m n-1)}{t + 4 m n-2} - t 2 m n \left(4 n^2-1\right)  \\
& + \left(2 m^2-m\right) \left(6 n^2-8 n^4\right)-  \left(2 m^2+m\right)2 n \left(4 n^3-8 n^2+3 n+1\right) \, . \label{cspII}
\end{aligned} 
\end{align}
Here $t$ is the level of $so(4 m n )$ affine algebra.
We find the relation between $c$ and $\ell$ as
\begin{align}
c = \frac{2 m \left(\ell^2+2 \ell m n-2 (\ell+1) (2 \ell+1) n^2\right)}{\ell+2 m n} \label{centerspII}
\end{align}
by removing the parameter $t$.

\subsection{OPEs among generators}
\label{sec:OPEs}

In the previous subsection, we have obtained the spin contents of the four types of W-algebras. 
In order to determine the algebras, we need the information of OPEs among generators as well.
In this subsection, we compute the OPEs by requiring their associativity for the simplest non-trivial examples with spin one and two generators as was done in \cite{Creutzig:2018pts}. For types $so(M (2 n+1))$ and $sp(2m (2 n +1))$, the minimal cases are with $n=1$, where the maximal spin is three. Therefore, we focus on the types $sp(2 M n)$ and $so(4 m n)$ with $n=1$.

\subsubsection*{\underline{Type $sp(2M n)$ with $n=1$}}

We consider the W-algebra with $so(M)$ symmetry obtained as the Hamiltonian reduction of $sp(2 M)$.
The spin one currents are the generators of $so(M)$ affine algebra.
The spin two currents are in the trivial and symmetric representations of $so(M)$.
In order to express the elements in the adjoint and symmetric representations of $so(M)$,
we utilize the $sl(M)$ generators $t^A_M$ expressed by $M \times M$ matrices.
We decompose $A = (a,\alpha)$ with $a=1,2,\ldots , M(M-1)/2$ and $\alpha = M(M-1)/2 +1 , \ldots , M^2 -1$ such that $t^a_M$ and $t^\alpha_M$ are the anti-symmetric and traceless symmetric matrices, respectively. 
We introduce the invariant tensors by
\begin{align}
	g^{AB} = \text{tr} (t^A_M t^B_M)  \, , \quad i f^{ABC} = \text{tr} ([t^A_M ,t^B_M] t^C_M) \, , \quad d^{ABC} = \text{tr} (\{t^A_M ,t^B_M\} t^C_M) 
	\label{tAB}
\end{align}
and use the convention $g^{AB} = \delta^{AB}$.
With this notation of indices, we use $J^a$ for the spin one currents, $T$ for the energy-momentum tensor, and $Q^\alpha$ for the charged spin two currents.

We would like to obtain OPEs among $J^a$, $T$, and $Q^\alpha$.
For the spin one currents $J^a$, we use the OPEs as
\begin{align}
J^a(z)J^b(0)&\sim\frac{\ell \kappa^{ab}}{z^2} + \frac{i f^{ab}_{~~c} J^c(0)}{z} \, . \label{JJOPE}
\end{align}
Here we have set $\kappa^{ab} = \frac12 g^{ab}$ for the normalization of level $\ell$ to be the conventional one. OPEs involving $T$ are
\begin{align}
T(z)T(0)\sim\frac{c/2}{z^4}+\frac{2T(0)}{z^2}+\frac{T ' (0)}{z} \, , \quad
T(z)J^a(0)\sim\frac{J^a(0)}{z^2}+\frac{{ J^a} ' (0)}{z} \, ,  \label{TTTJOPE}
\end{align}
where  $A'(0) = \frac{d}{dz} A(z)|_{z = 0}$.
Here $c$ is the central charge of the algebra.
As in \eqref{centersoII}, there would be a relation between $\ell$ and $c$,
but we do not require the relation for a while.
The currents $Q^\alpha$ have spin two and transform as in the symmetric representation of $so(M)$.
Using \eqref{tAB}, we require the OPEs 
\begin{align}
T(z)Q^\alpha(0)\sim \frac{2Q^\alpha(0)}{z^2} + \frac{{Q^\alpha}' (0)}{z} \, , \quad
J^a(z)Q^\alpha(0)\sim \frac{i f^{a\alpha}_{~~\beta}Q^\beta(0)}{z}\, . \label{TQJQOPE}
\end{align}
Non-trivial OPEs left are $Q^\alpha (z) Q^\beta(0)$, and we obtain them by requiring associativity of the OPEs.

The operator products $Q^\alpha (z) Q^\beta(0)$ would generate (composite) operators up to spin three, so we classify possible operators primary w.r.t.~the Virasoro generator $T$.
In order to make composite operators well defined, we adopt the normal ordering prescription
\begin{align}
\begin{aligned}
&\left( AB \right)(z)=\frac{1}{2\pi i}\oint\frac{dw}{w-z}A(w)B(z) \, , \\
&(A_1 A_2 \cdots A_p) = (A_1 \cdots (A_{p-2} (A_{p-1} A_p)) \cdots) \, . 
\end{aligned} \label{normalorder}
\end{align}
Spin one primary operators are $J^a$. For spin two, there are 
\begin{align}
\left[J^{(a }J^{b)}\right] =(J^{(a}J^{b)})-\frac{2 \ell}{c} \kappa^{ab}T  \label{spin2comp}
\end{align}
along with $Q^\alpha$.
Here the brackets $(a_1,a_2,\ldots, a_p)$ and $[a_1,a_2,\ldots,a_p]$ mean the symmetric and anti-symmetric indices, respectively, with the factor $1/(p!)$.
For spin three, composite primary operators are 
\begin{align}
	\begin{aligned}
		&\left[J^{(a}J^bJ^{d)}\right] =(J^{(a}J^bJ^{d)})+\frac{3 \ell}{c+2}\left \{ \kappa^{(ab} {J^{d)}}'' -2 \kappa^{(ab}(TJ^{d)})\right\} \, , \\
		&\left[{J^{[a}}' J^{b]}\right]=\left({J^{[a}}' J^{b]} \right)-\frac{i}{6(c+2)}f^{ab}_{~~d}\left \{ (c-4){J^d} '' +12(TJ^d)\right \} \, ,\\
        &\left[J^a Q^\alpha \right] =(J^a Q^\alpha) - \frac{i}{4}f^{a\alpha }_{~~\beta} {Q^\beta } ' \, .	\end{aligned} \label{spin3comp}
\end{align}

We expand the operator products $Q^\alpha (z) Q^\beta (0)$ in terms of these (composite) primary operators with several indices. In order to express the coefficients, we need invariant tensors consisting of traces of $sl(M)$ generators as classified in  \cite{Creutzig:2018pts}.
With two and three indices, we use the tensors in \eqref{tAB}.
The invariant tensors with four indices are
\begin{align}
	    & d^{ABCD}_{4SS0} \equiv g^{AB} g^{CD} \, , \quad
		 d^{ABCD}_{4SS1} \equiv 4 \text{tr}(t^{(A}t^{B)}t^{(C}t^{D)}) \, ,\quad 
		 d^{AB}_{4SS2,CD} \equiv 4 \text{tr}(t^{(A}t_{(C}t^{B)}t_{D)})\, , \nonumber \\
		& d^{AB}_{4SS3,CD} \equiv \delta^A_{~(C} \delta^B_{~D)} \, , \quad
		 d^{ABCD}_{4SA}\equiv  d^{ABE} f_{E}^{~~CD}  \, , \quad
		 d^{ABCD}_{4AA1}\equiv  f^{ABE} f_{E}^{~~CD} \, , \label{4tensors}\\& d^{AB}_{4AA2,CD}\equiv 4 \text{tr}(t^{[A}t_{[C}t^{B]}t_{D]}) \, , \quad  
		 d^{AB}_{4AA3,CD} \equiv \delta^A_{~[C} \delta^B_{~D]} \, . \nonumber  
\end{align}
In addition, we also need invariant tensors with five indices as
\begin{align}
\begin{aligned}
		&d^{ABCDE}_{50}\equiv  f^{AB(C} g^{DE)}\, ,\quad 
		d^{ABCDE}_{51}\equiv 12 \text{tr}(t^{[A}t^{B]}t^{(C}t^D t^{E)})\, ,\\ 
		&d^{AB}_{52,CDE}\equiv 12 \text{tr}(t^{[A}t_{(C}t^{B]}t_{D} t_{E)})\, , \quad
		d^{AB}_{53,CDE}\equiv  \delta^{[A}_{~(C} d^{B]}_{~DE)}\, .
\end{aligned} \label{5tensors}
\end{align}

With the preparations, the ansatz for the OPEs of $Q^\alpha (z) Q^\beta (0) $ can be written as
\begin{align}
		&Q^\alpha (z)Q^\beta (0)\sim \frac{c_1}{c}g^{\alpha \beta}\left(\frac{c/2}{z^4} + \frac{2 T(0)}{z^2} + \frac{ T'(0)}{z} \right) \nonumber \\
		&+c_2 i f^{\alpha \beta}_{~~a}\left( \frac{J^a(0)}{z^3} + \frac{\frac12  {J^a}' (0)}{z^2} + \frac{\left(\frac{1}{6} - \frac{2}{c+2} \right) {J^a}' (0)+ \frac{4}{c+2} (TJ^a)(0)}{z} \right) \nonumber\\
		&+ ( c_{31} d_{4SS0}^{\alpha \beta a b}+ c_{32} d_{4SS1}^{\alpha \beta a b} + c_{33} d_{4SS2}^{\alpha \beta a b} + c_{34} d_{4SS3}^{\alpha \beta a b}) \left( \frac{ \left[J_{(a}J_{b)}\right](0)}{z^2} + \frac{ \frac12  \left[J_{(a}J_{b)} \right] ' (0)}{z} \right) \label{QQansatz} \\
		&+ (c_{41} d_{4AA1}^{\alpha \beta a b} + c_{42} d_{4AA2}^{\alpha \beta a b} +c_{43} d_{4AA3}^{\alpha \beta a b} ) \frac{\left[{J_{[a}}' J_{b]}\right](0)}{z}\nonumber \\
		&+ (c_{51}  d_{50}^{\alpha \beta abc} + c_{52} d_{51}^{\alpha \beta abc} + c_{53} d_{52}^{\alpha \beta abc} +c_{54} d_{53}^{\alpha \beta abc} ) \frac{\left[J_{(a}J_b J_{c)}\right] (0)}{z} \nonumber\\
		&+ c_6 d^{ab}_{~~\alpha} \left( \frac{2 Q^\alpha (0)}{z^2} + \frac{\partial Q^\alpha (0)}{z}  \right ) 
		+ (c_{71} d_{4AA1}^{\alpha \beta a \gamma} + c_{72} d_{4AA2}^{\alpha \beta a \gamma} + c_{73} d_{4SA}^{\alpha \beta a \gamma} + 2 c_{74} d_{4AA3}^{\alpha \beta a \gamma}) \frac{\left[J_{a}Q_{\gamma}\right] (0)}{z} \, . \nonumber
\end{align}
Requiring associativity of the OPEs, we can uniquely fix the coefficients of our ansatz in terms of $\ell$ as%
\footnote{For the computation, we use the Mathematica package ``OPEdefs'' \cite{Thielemans:1991uw}.
We have obtained the results for $M=3,4,5,6$ and propose that the expressions are true also for $M>6$.
}
\begin{align}
&c_2=\frac{c_1}{\ell} \, ,\quad
c_{31}=\frac{2 c_1 \left(\ell \left(5 \ell M+4 (\ell-2) \ell+2 M^2\right)-16 M\right)}{(\ell-4) (\ell-1) \ell M (2 \ell+M-4) (3 \ell+2 M-4)} \, , \nonumber \\
&c_{32}=\frac{ c_1 (\ell-2)}{(\ell-1) \ell (2 \ell+M-4)} ~, \quad c_{33}=\frac{ c_1}{2 \ell-2 \ell^2} \, , \quad c_{34}=0 \, ,  \nonumber \\
&c_{41}=\frac{ c_1 \left(\ell^3 M+\ell^2 (M-4)^2+\ell (-6 (M-8) M-32)+16 (M-4) M\right)}{(\ell-4) (\ell-1) \ell M (2 \ell+M-4) (3 \ell+2 M-4)} \,  , \nonumber \\
&c_{42}=0 \, , \quad c_{43}=0 \, ,\quad
c_{51}=\frac{8 c_1 (\ell+M)}{(\ell-4) (\ell-1) \ell M (3 \ell+2 M-4)}\, ,   \\
&c_{52}=\frac{2 c_1 (\ell-4)}{3 (\ell-1) \ell (2 \ell+M-4) (3 \ell+2 M-4)} \, ,\quad
c_{53}=-\frac{2 c_1 }{3 (\ell-1) \ell (2 \ell+M-4)}\, ,  \nonumber  \\ &c_{54}=0 \, ,\quad
c_6= \pm \frac{2 i \sqrt{c_1} (\ell-2) (\ell+M)}{\sqrt{(\ell-4) (\ell-1) \ell (2 \ell+M-4) (3 \ell+2 M-4)}} \, ,  \nonumber \\
&c_{71}=0 \, , \quad c_{72}=\frac{c_6}{\ell-2} \, , \quad
c_{73}=\frac{i c_6 (\ell-4)}{(\ell-2) (\ell+M)}\, , \quad c_{74}=0  \nonumber 
\end{align}
up to an overall normalization $\pm\sqrt{c_1}$ for the spin two currents $Q^\alpha$.
The associativity also fixes the central charge as
\begin{align}
c=\frac{M \left(-3 \ell ^2+2 \ell (M+6)-8\right)}{2 (\ell+2 M)} \, ,
\end{align}
which reproduces \eqref{centersoII} with $n=1$.

\subsubsection*{\underline{Type $so(4 m n)$ with $n=1$}}

We then move to the W-algebra with $sp(2m)$ symmetry obtained as the Hamiltonian reduction of $so(4 m)$.
The spin one currents generate $sp(2m)$ affine algebra.
A spin two current is the energy-momentum tensor, and the others are in the anti-symmetric representation of $sp(2m)$. In order to express the adjoint and anti-symmetric representations simultaneously, we again use the $sl(2m)$ generators $t^A_{2m}$ expressed by $2 m \times 2 m$ matrices.
In this case, we decompose the indices as $A =(a,\alpha)$ with $a= 1 ,2 \ldots , 2m^2 + m$ and $\alpha =2 m^2 + m + 1, \ldots , 4 m^2 -1$ such that
\begin{align}
\Omega ^{-1}_{2m} ( t^a_{2m} ) ^\text{tr}\Omega_{2m} = - t^a_{2m} \, , \quad \Omega _{2m} ^{-1} ( t^\alpha _{2m}) ^\text{tr}\Omega_{2m} = t^\alpha _{2m} \, , \quad \Omega^\text{tr} _{2m} = - \Omega _{2m} \, . \label{conditionofsp}
\end{align}
We introduce the invariant tensors as in \eqref{tAB} with $M=2m$.
In particular, we raise or lower indices by $g^{AB}$ or its inverse.
With these indices, the spin one and two currents are denoted as $J^a$, $T$, and $Q^\alpha$.

For OPEs among $J^a$ and $T$, we use \eqref{JJOPE} and \eqref{TTTJOPE}.
Here we set $\kappa^{ab} = g^{ab}$ for the convention of $\ell$ to be as usual.
Again, we do not specify the relation between  $\ell$ and $c$ for a while.
We set $Q^\alpha$ such that the OPEs \eqref{TQJQOPE} are satisfied.
For the ansatz of the OPEs of $Q^\alpha (z) Q^\beta (0)$, we use the same form as in \eqref{QQansatz} with the same expressions of composite operators \eqref{spin2comp}, \eqref{spin3comp}.
Notice that the explicit expressions differ since the invariant tensors \eqref{tAB}, \eqref{4tensors}, \eqref{5tensors} are constructed from the different set of generators $t^A _{2m}= (t^a _{2m} , t^\alpha _{2m})$.

Requiring associativity of the OPEs, we can uniquely fix the coefficients of our ansatz as%
\footnote{We have obtained the expressions for $m=2,3,4$ and propose that they are true also for $m>4$.}
\begin{align}
&c_2=\frac{c_1}{2 \ell}\, ,\quad
c_{31}= \frac{c_1 (\ell (\ell (\ell ((m-9) m+26)+(m-9) m+56)+12 m)-24)}{12 \ell (\ell+2) (2 \ell+1) (2 \ell+m+2) (3 \ell+2 m+2)}\, , \nonumber \\
&c_{32}=\frac{c_1 (\ell+1)}{2 \ell (2 \ell+1) (2 \ell+m+2)}~, \quad c_{33}=-\frac{c_1}{4 \ell (2 \ell+1)}~, \quad c_{34}=0\, , \nonumber\\
&c_{41}=\frac{c_1 \left(\ell \left(6 \ell^2+\ell ((m-3) m+50)+m (m+9)+98\right)+24 (m+2)\right)}{12 \ell (\ell+2) (2 \ell+1) (2 \ell+m+2) (3 \ell+2 m+2)}\, , \nonumber\\
&c_{42}=0\, , \quad 
c_{43}=0\, ,\quad
c_{51}=\frac{c_1 (\ell+m)}{\ell (\ell+2) (2 \ell+1) m (3 \ell+2 m+2)}\, ,\\
&c_{52}=\frac{c_1 (\ell+2)}{6 \ell (2 \ell+1) (2 \ell+m+2) (3 \ell+2 m+2)}\, ,\quad
c_{53}=-\frac{c_1}{6 \ell (2 \ell+1) (2 \ell+m+2)}\, ,\nonumber \\
&c_{54}=0\, ,\quad
c_6=\pm\frac{2 i \sqrt{c_1} (\ell+1) (\ell+m)}{\sqrt{\ell (\ell+2) (2 \ell+1) (2 \ell+m+2) (3 \ell+2m+2))}}\, , \nonumber\\
&c_{71}=0\, , \quad c_{72}=-\frac{c_6}{2 (\ell+1)}\, , \quad 
c_{73}=\frac{i c_6 (\ell+2)}{2 (\ell+1) (\ell+m)}\, , \quad   c_{74}=0  \nonumber
\end{align}
up to the overall normalization $\pm \sqrt{c_1}$ of $Q^\alpha$.
The central charge is also determined as
\begin{align}
c=-\frac{2 m \left(3 \ell^2-2 \ell (m-3)+2\right)}{\ell +2 m} \, ,
\end{align}
which is the same as the expression of  \eqref{centerspII} with $n=1$.

\subsubsection*{\underline{Structure of null vectors}}

In subsection \ref{sec:hsa}, we have explained that there are two types of restricted matrix extensions of  $hs[\lambda]$, which may be called as $hs_{so(M)}[\lambda]$ and $hs_{sp(2m)}[\lambda]$.
Moreover, we have derived that $hs_{so(M)}[\lambda]$ and $hs_{sp(2m)}[\lambda]$ reduce to $so$ or $sp$ type Lie algebras at positive integer $\lambda$ by quotienting ideals. From the Hamiltonian reduction of $hs_{so(M)}[\lambda]$ or $hs_{sp(2m)}[\lambda]$, we can obtain a rectangular W-algebra with $so(M)$ or $sp(2m)$ symmetry, which may be denoted as W$_{so(M)}[\lambda]$ or W$_{sp(2m)}[\lambda]$. It is natural to expect that our rectangular W-algebras can be obtained from W$_{so(M)}[\lambda]$ and W$_{sp(2m)}[\lambda]$ at positive integer $\lambda$ by quotienting ideals.
In particular, the spin three currents in these W-algebras are assumed to be decoupled at $\lambda =2$, and this assumption leads to the algebras obtained in this subsection.
As claimed below, our W-algebras are supposed to be reproduced from coset models with labels $(k,N,M)$ or $(k,N,m)$. From the correspondences, we expect that other truncations could occur with integer $k,N$, since coset models with positive integer parameters will have extra null vectors. 
However, it is an important open problem to reveal the structure of null vectors of W$_{so(M)}[\lambda]$ and W$_{sp(2m)}[\lambda]$.
In a coset model similar to ours, these two types of truncation were discussed in subsection 3.4 of \cite{Beccaria:2014jra}, see also \cite{Gaberdiel:2013vva,Gaberdiel:2014yla,Creutzig:2019qos}.

\subsection{Comparison with dual coset models}
\label{sec:map}

As in \cite{Creutzig:2018pts}, we would like to relate the W-algebras with $so(M)$ or $sp(2m)$ symmetry to the symmetry algebras of some coset models. The corresponding coset models should be the holographic duals of the restricted matrix extensions of 3d Prokushkin-Vasiliev theory. 
The symmetry algebras become quite simple at the limit when the parameter $\lambda$ in the gauge algebras becomes $\lambda \to 0$ or $\lambda \to 1$. At the limit, we can realize the rectangular W-algebras as orbifolds of free bosons and fermions \cite{Creutzig:2012sf, Creutzig:2014lsa}, and we can easily obtain the cosets which reduce to the same free systems at the large level limit \cite{Bakas:1990xu,Odake:1990rr}, see also \cite{Creutzig:2013tja,Eberhardt:2018plx}.
Our proposal on the dual coset CFTs can be found in table \ref{fig:proposal}.%
\footnote{See appendix \ref{sec:alternative} for alternative proposals on the dual coset CFTs. 
	It is an interesting open problem to examine the relation between two proposals on the dual cosets.} 
A difficult issue is on the maps of parameters including quantum corrections.
In this subsection, we achieve this by making use of the levels $\ell$ of $so(M)$ or $sp(2m)$ affine symmetries  and the central charges $c$ of the algebras.

\subsubsection*{\underline{Type $so(M(2 n +1))$}}

We start from the W-algebra with $so(M)$ symmetry realized as the Hamiltonian reduction of $so(M(2 n+ 1))$.
As a natural guess,  we propose that the corresponding coset is given by
\begin{align}
\frac{so(N + M)_k}{so(N)_k} \, . \label{cosetsoI}
\end{align}
The coset has the $so(M)$ affine symmetry and the level is $k$.
The central charge is computed as
\begin{align}
c = \frac{k ( N + M ) (N + M  - 1)}{2 (k + N + M - 2 )} - \frac{k N (N -1)}{2(k + N -2 )} \, , \label{centralsoI}
\end{align}
where the dual Coxeter number of $so(L)$ is given by $h^\vee = L - 2$.
At the large $k$ limit, the coset reduces to $M N$ real bosons with $so(N)$ invariance  in addition to the $so(M)$ currents.
Compared with the level \eqref{lsoI} and the central charge \eqref{csoI}, the correspondence is realized with $\ell = k$ and $\lambda = 2 n + 1$, where the 't Hooft parameter is given by
\begin{align}
\lambda = \frac{k - 2}{k + N  - 2} \, , \quad \lambda = - \frac{k - 2}{k+M+N-2} \, . \label{thooftsoI}
\end{align}
The existence of two 't Hooft parameters implies a duality relation of the coset \eqref{cosetsoI}.
The same kind of duality exists in the W-algebra with $su(M)$ symmetry, and it was analyzed in \cite{Creutzig:2018pts}  to some extent.

\subsubsection*{\underline{Type $sp(2 M n)$}}

We then consider the W-algebra with $so(M)$ symmetry obtained from the Hamiltonian reduction of $sp(2 M n)$.
We propose that the algebra can be realized by the coset
\begin{align}
	\frac{osp(M| 2 N )_k}{sp(2N)_k} \, . \label{cosetsoII}
\end{align}
The symmetry algebra of the coset includes $so(M)$ affine Lie algebra with level $- 2 k$ and moreover it has $m(m+1)/2$ generators at conformal weights $2, 4, \dots, 2N$ and $m(m-1)/2$ generators at conformal weights $1, 3, \dots, 2N+1$ by Theorem 3.15 of \cite{Creutzig:2016xos} together with the theory of cosets of  \cite{Creutzig:2012sf, Creutzig:2014lsa}.
The central charge of the coset is
\begin{align}
	c = \frac{k (M (M-1)/2 + N (2 N +1)  - 2 N M )}{k + N - M/2 +1} - \frac{k N (2 N +1)}{k + N + 1} \, .
	\label{centralsoII}
\end{align}
For the dual Coxeter number  of $osp(M| 2 N)$,  we have used $h^\vee = N - M/2 + 1$, see, e.g., table 5 of \cite{Kac_2000}. The dual Coxeter number of $sp(2 L)$ is $h^\vee = L + 1$.
At large $k$, the coset reduces to $2 M N$ Majorana fermions with $sp(2N)$ invariance as well as the $so(M)$ currents.
The comparison with \eqref{lsoII} and \eqref{csoII} leads to $\ell = - 2k$ and $\lambda = 2n$, where the 't Hooft parameter $\lambda$ is defined as
\begin{align}
	\lambda = \frac{k}{k + N +1} \, , \quad \lambda = - \frac{k}{k + N  - M/2 + 1} \, .
	\label{tHooftsp}
\end{align}
There are two correspondences and this implies a duality of the coset \eqref{cosetsoII}.

\subsubsection*{\underline{Type $sp(2 m (2 n+1))$}}

We move to the W-algebra with $sp(2m)$ symmetry realized as the Hamiltonian reduction of $sp(2 m (2 n+1))$.
We propose that the corresponding coset is 
\begin{align}
\frac{osp(N| 2 m)_k}{so(N)_k} \, . \label{cosetspI}
\end{align}
The level of the $sp(2 m)$ affine algebra is $-k/2$, and
the central charge of the coset is
\begin{align}
c = \frac{k (N (N-1)/2 + m (2 m +1) - 2 N m )}{k + N - 2 m  - 2} - \frac{k N (N - 1)}{2 ( k + N -2 )} \, . \label{centralspI}
\end{align}
For the dual Coxeter number  of $osp(N| 2 m)$,  we have used $h^\vee = N - 2m -2$ in this case.
At the large $k$ limit, the coset reduces to $2 m N$ Majorana fermions with $so(N)$ invariance in addition to the $sp(2m)$ currents.
Compared with the expressions \eqref{lspI} and \eqref{cspI}, we find that the correspondence happens at $\ell = - k/2, \lambda = 2 n + 1$ with
\begin{align}
\lambda = \frac{k - 2}{k + N  - 2} \, , \quad \lambda = - \frac{k -2}{ k  + N - 2 m - 2} \, .  \label{thooftspI}
\end{align}
Again, the possibility of two choices implies  a duality of the coset \eqref{cosetspI}.

\subsubsection*{\underline{Type $so(4 m n)$}}

Finally, we examine the W-algebra with $sp(2m)$ symmetry obtained by the Hamiltonian reduction of $so(4 m n)$.
The symmetry algebra is proposed to be realized by
\begin{align}
	\frac{sp( 2  N + 2 m)_k}{sp(2 N)_k}  \, . \label{cosetspII}
\end{align}
The coset includes $sp(2 m)$ affine symmetry with level $k$.
The central charge is computed as 
\begin{align}
	c = \frac{k ( N + m ) (2 N + 2 m  +1)}{k + N + m + 1} - \frac{k N (2 N +1)}{k + N + 1} \, . \label{centralspII}
\end{align}
At the large level limit, the coset reduces to $4 m N$ real bosons with $sp(2N)$ invariance as well as the $sp(2m)$ currents.
Compared with \eqref{lspII} and \eqref{cspII}, we find that the correspondence is realized at $\ell = k, \lambda = 2n$ with
\begin{align}
 \lambda = \frac{k}{k+N+1} \, , \quad \lambda = -\frac{k}{k+N+m +1} \, .  \label{thooftspII}
\end{align}
There should be a duality of the coset \eqref{cosetspII} as in the other cases.

\subsection{Symmetry of dual coset models}
\label{sec:cosetalgabra}

In subsection \ref{sec:OPEs}, we have obtained OPEs for two simple examples  with generators of spin up to two.
In this subsection, we construct the generators of the restricted rectangular W-algebras in terms of dual cosets and check that the OPEs obtained above can be reproduced.

\subsubsection*{\underline{Type $sp(2M n)$ with $n=1$}}

We have proposed above that the W-algebra with $so(M)$ symmetry obtained from $sp(2 M)$ can be realized as the symmetry algebra of  \eqref{cosetsoII} at $k = - 2 (N +1)$ (or $k = -2 /3 (N - M/2 +1)$).
The symmetry generators are constructed from the currents in the numerator  and should be regular w.r.t.~the currents in the denominator.
In order to write down the $osp(M|2N)$ currents, we fist introduce the generators of 
the Lie superalgebra $osp(M|2N)$.
We use $(M + 2 N) \times (M + 2 N)$ even supermatrices
\begin{align} 
M_{M|2N} =
\begin{pmatrix}
A_M & B_{M,2N} \\
C_{2N,M} & D_{2N} 
\end{pmatrix} 
\end{align}
and require the conditions
\begin{align}
A^\text{tr}_M = - A_M \, , \quad \Omega^{-1}_{2N} D^\text{tr}_{2N} \Omega_{2N} = - D_{2N} \, , \quad \Omega^{-1}_{2N} B^\text{tr}_{M,2N} = C_{2N,M} \, , \quad \Omega ^\text{tr}_{2N} = - \Omega_{2N} \, , 
\end{align}
see \eqref{ospcond1} and \eqref{ospcond2}.
As generators, we use
\begin{align}
t^a _{M|2N}=
\begin{pmatrix}
t^a_M & 0 \\
0 & 0 
\end{pmatrix} \, , \quad
t^p_{M|2N} =
\begin{pmatrix}
0 & 0 \\
0 & t^p_{2N}
\end{pmatrix} \, , \quad
t^i _{M|2N}=  \begin{pmatrix}
0 & t^i_{M , 2N} \\
\Omega^{-1}_{2N} ( t^i_{M , 2N} )^\text{tr}  & 0
\end{pmatrix} \, . \label{ospgenerators}
\end{align}
Here $t^a_M$ and $t^p_{2N}$ are $so(M)$ and $sp(2N)$ generators, respectively.
For $t^{i}_{M , 2N}$ we use a $M \times 2 N$ matrix where only one element is non-zero and equal to one. 
We also need 
\begin{align}
t^\alpha _{M|2N} =
\begin{pmatrix}
t^\alpha_M & 0 \\
0 & 0 
\end{pmatrix} \, , \label{ospgenerators2}
\end{align}
where $t^\alpha_M$ are generators for the symmetric representation of $so(M)$.
Denoting $t^P _{M|2N}= (t^a_{M|2N}, t^\alpha_{M|2N}, t^p_{M|2N},t^i_{M|2N} )$, we introduce the metric
\begin{align}
g^{PQ} = \text{str} (t^P_{M|2N} t^Q_{M|2N})  \label{tPQ0}
\end{align}
and the invariant tensors 
\begin{align}
 i f^{PQR} = \text{str} ([t^P _{M|2N}, t^Q_{M|2N}] t^R_{M|2N}) \, , \quad 
d^{PQR} = \text{str} (\{t^P_{M|2N} ,t^Q _{M|2N}\} t^R_{M|2N}) \, . \label{tPQ}
\end{align}
Here str is defined in \eqref{str}.
For $t^a_{M},t^\alpha_{M}$, we use the same generators as in subsection \ref{sec:OPEs}.
In particular, the normalizations are such that $g^{ab} = \delta^{ab}$ and $g^{\alpha \beta} = \delta^{\alpha \beta}$.
The osp$(M | 2 N)$ currents include bosonic ones $J^a,J^p$ and fermionic ones $\tilde J^i$.
The non-trivial OPEs are
\begin{align}
\begin{aligned}
&J^a (z) J^b (0) \sim \frac{- k  \delta^{ab} }{z^2} + \frac{i f^{ab}_{~~c} J^c (0)}{z} \, , \quad
J^p (z) J^q (0) \sim \frac{- k g^{pq}}{z^2} + \frac{i f^{pq}_{~~~r} J^r (0)}{z} \, , \\
&\tilde J^i (z) \tilde J^j (0) \sim \frac{- k g^{ij}}{z^2} + \frac{d^{ij}_{~~a} J^a (0)+ d^{ij}_{~~p} J^p (0)}{z} \, ,\\ & J^a (z) \tilde J^i (0) \sim \frac{i f^{ai}_{~~j} \tilde J^j (0)}{z} \, , \quad
J^p (z) \tilde J^i (0) \sim \frac{i f^{pi}_{~~j} \tilde J^j (0)}{z} \, .
\end{aligned} \label{ospOPE}
\end{align}

With these preparations, we write down the generators of the  W-algebra with $so(M)$ symmetry up to spin two. For the spin one generators, we have $J^a$ with level $-2 k$.
There are spin two generators, such as, the energy-momentum tensor $T$ and the charged spin two currents. 
The energy-momentum tensor is obtained by the standard coset construction \cite{Goddard:1984vk}.
The charged spin two generators are given by linear combinations as
\begin{align}
\hat  Q^\alpha = d^{\alpha }_{~ij} (\tilde J^i \tilde J^j) + \frac{2 N   }{4 k - M + 4 } d^{\alpha}_{~ab} (J^{a} J^{b})  \, . \label{tQso}
\end{align}
Here the relative coefficients are fixed so as to satisfy the OPEs \eqref{TQJQOPE}. 
Using the OPEs in \eqref{ospOPE}, we can explicitly evaluate the OPEs of $\hat Q^\alpha (z) \hat Q^\beta (0)$. In fact, we find%
\footnote{We have checked this for $M=3,4,5,6$ and $N=1,2,3$.}
\begin{align}
	\hat Q^\alpha (z) \hat Q^\beta (0)\sim Q^\alpha (z)Q^\beta (0) + \frac{if^{\alpha \beta }_{~~\gamma }P^\gamma (0)}{z}
	\label{KKQQ}
\end{align}
with extra spin three currents $P^\alpha$ at $k = - 2 (N+1)$.
Since we can show
$
	P^\alpha (z)P^\beta (0)\sim \mathcal{O}(z^{-5}) 
$,
the new operators $P^\alpha$ are regarded as null vectors.
Therefore, we can decouple $P^\alpha$ from the rest by setting $P^\alpha=0$ and identify the OPEs of $\hat Q^\alpha$ and $Q^\alpha$ by \eqref{KKQQ}.%
\footnote{The spin three operators $P^\alpha$ are identified as null vectors of W$_{so(M)}[\lambda]$ appearing at $\lambda = 2$ discussed at the end of subsection \ref{sec:OPEs}. A similar statement holds also for the W-algebra with $sp(2m)$ symmetry analyzed below.}

\subsubsection*{\underline{Type $so(4 m n)$ with $n=1$}}

We consider the W-algebra with $sp(2m)$ symmetry obtained from $so(4m)$.
In the previous subsection, we have proposed that the algebra can be realized as the symmetry algebra of the coset \eqref{cosetspII} at $k = - 2 (N +1)$ (or $k = - 2/3 (N + m +1)$).
In order to express the $sp(2 m + 2 N)$ currents, we 
decompose the generators of $sp( 2m + 2 N)$ in the following way.
We use $(2m + 2 N) \times (2m + 2 N)$ matrices
\begin{align}
M_{2m+2N} =
\begin{pmatrix}
A_{2m} & B_{2m,2N} \\
C_{2N ,2m} & D_{2N} 
\end{pmatrix}
\end{align}
and realize the condition of  $sp( 2m + 2 N)$  as
\begin{align}
\Omega_{2 m}^{-1 } A^\text{tr}_{2m} \Omega_{2m} = -A_{2m} \, , \quad
\Omega_{2 N}^{-1 } D^\text{tr}_{2N} \Omega_{2N} = -D_{2N} \, , \quad
\Omega_{2N}^{-1} B^\text{tr}_{2m ,2N} \Omega_{2m} = - C_{2N,2m} \, .
\end{align}
Here $\Omega_{2m}$ and $\Omega_{2N}$ are symplectic forms satisfying
\begin{align}
\Omega^\text{tr}_{2m} = - \Omega_{2m} \, , \quad 
\Omega^\text{tr}_{2N} = - \Omega_{2N} \, .
\end{align}
For generators, we use
\begin{align}
\begin{aligned}
&t^a _{2 m + 2 N}=
\begin{pmatrix}
t^a_{2m} & 0 \\
0 & 0 
\end{pmatrix} \, , \quad
t^p _{2 m + 2 N}= 
\begin{pmatrix}
0 & 0 \\
0 & t^p_{2N} 
\end{pmatrix} \, , 
\\
&t^i _{2 m + 2 N}=
\begin{pmatrix}
0 & t^i_{2 m , 2N} \\ - 
\Omega_{2N}^{-1} (t^{i}_{2 m , 2 N})^\text{tr} \Omega_{2m} & 0 
\end{pmatrix} \, , 
\end{aligned}
\label{spgenerators}
\end{align}
where $t^a_{2m}$ and $t^p_{2 N}$ are the generators of $sp(2m)$ and $sp(2N)$ subalgebras, respectively.
Moreover,  each $t^i_{2m , 2N}$ has only one non-zero element, which is equal to one.
We also need
\begin{align}
t^\alpha _{2 m + 2 N} =
\begin{pmatrix}
t^\alpha_{2m} & 0 \\
0 & 0 
\end{pmatrix} \, ,  \label{spgenerators2}
\end{align}
where $t^\alpha_{2m}$ are generators for the anti-symmetric representation of $sp(2m)$.
With $t^P _{2 m + 2 N}= (t^a_{2 m + 2 N}  , t^\alpha_{2 m + 2 N} ,t^p _{2 m + 2 N},t^i _{2 m + 2 N})$, we introduce the metric as in \eqref{tPQ0} and the invariant tensors as in \eqref{tPQ}.
We use the same expressions for $t^a_{2m},t^\alpha_{2m}$ as in subsection \ref{sec:OPEs} in order to make comparison easier.
The OPEs of $sp(2m+2N)$ currents are given by
\begin{align}
J^X (z) J^Y (0) \sim \frac{k g^{XY}}{z^2} + \frac{i f^{XY}_{~~~Z} J^Z (0)}{z} 
\label{spOPE}
\end{align}
with $J^X= (J^a  , J^p  , J ^i )$.

We construct the generators of the rectangular W-algebra up to spin two in the coset language.
Spin one generators are $J^a$ with level $k$. As a spin two current, the energy-momentum tensor $T$ can be constructed in the standard way \cite{Goddard:1984vk}. The other spin two currents are given by
\begin{align}
\hat Q^\alpha = d^{\alpha}_{~ij} (J^i J^j) - \frac{N}{2 k + m + 2 } d^{\alpha}_{~ab} ( J^a J^b) \, , \label{tQsp}
\end{align}
which satisfy the OPEs \eqref{TQJQOPE}.
Using \eqref{spOPE}, the OPEs of $\hat Q^\alpha (z) \hat Q^\beta (0)$ can be evaluated as \eqref{KKQQ}
with extra spin three currents $P^\alpha$ at $k = - 2 (N+1)$.%
\footnote{We have checked this with $m=2,3,4$ and $N=1,2,3,4$ except for $m=N+1$, where the central charge diverges.}
We can show that
$P^\alpha $ have zero norm as
$
P^\alpha (z)P^\beta (0)\sim \mathcal{O}(z^{-5})
$. Thus, we decouple $P^\alpha $ and identify the OPEs of $\hat Q^\alpha$ and $Q^\alpha$ by \eqref{KKQQ}.

\section{$\mathcal{N}=1$ rectangular W-algebras}
\label{sec:super}

The gauge algebras for Chern-Simons supergravities have been analyzed in subsection \ref{sec:hssa}, and it was found that there are four types of restricted matrix extensions, see table \ref{fig:hsa}.
We  assign the asymptotic AdS boundary condition as in \eqref{AdScond} for higher spin supergravity theories. Under this condition, the asymptotic symmetries are again given by the Hamiltonian reductions of superalgebras with the $osp(1|2)$ embeddings \cite{Creutzig:2011fe,Hanaki:2012yf,Henneaux:2012ny}.
In this way, we obtain four types of $\mathcal{N}=1$ rectangular W-algebras. 
In the next subsection, we explain their basic properties, such as, spin contents, central charges, and the levels of affine symmetries. In subsection \ref{sec:superOPEs}, we show that the associativity gives strong constraints on OPEs  for simple examples with generators of spin up to two.
In subsection \ref{sec:supermap}, we argue that the $\mathcal{N}=1$ W-algebras could be realized as the symmetry algebras of coset models listed in \ref{fig:proposal}. In subsection \ref{sec:superdualcosets}, we explicitly construct generators up to spin two in terms of the cosets and check that the OPEs can be reproduced for several examples.

\subsection{Basics properties of the superalgebras}
\label{sec:superbasics}

In this subsection, we study the basic properties of the four types of  $\mathcal{N}=1$ W-algebras obtained by the Hamiltonian reductions of the superalgebras listed in table \ref{fig:hsa}.
Specifically, we examine the spin contents of these superalgebras and write down the levels $\ell_1,\ell_2$ of two $so(M)$ or $sp(2m)$ affine subalgebras and the central charges $c$ of the algebras, see appendix \ref{sec:candks} for the details of computations. 

\subsubsection*{\underline{Type $osp(M (2 n+1)| 2 M n)$}}

We start from the case with $\mathfrak{g} = osp(M (2 n+1)| 2 M n)$.
We decompose the superalgebra as in \eqref{tsalgebra0}.
Here $\mathbbm{1}_M \otimes B^\sigma_{2n+1|2n}$ generates $osp(2 n+1 |2n)$, and the supergravity sector is identified with  the principally embedded $osp(1|2)$ in the $osp(2 n+1 |2n)$.
With the $sl(2) \subset osp(1|2)$, the superalgebra $osp(M (2 n+1)| 2 M n)$ can be decomposed as
\begin{align}
\begin{aligned}
osp(M(2 n+1)| 2 M n ) 
\simeq \, &  M (M-1)  \left(\bigoplus_{s=0}^{n-1}  \rho_{2s} \right)   \oplus M^2 \left(\bigoplus_{s=1}^{2n} \rho_{s-1/2} \right)  \\
&\oplus M ( M+1) \left(\bigoplus_{s=1}^{n} \rho_{2s-1} \right)    \oplus \frac{M(M-1)}{2} \rho_{2n}\, .
\end{aligned}
\end{align}
As in the bosonic case, after the Hamiltonian reduction, only one element in each $\rho_s$ or $\rho_{s-1/2}$ survives. Therefore, the W-algebra after the reduction includes $M(M-1)$ currents with spin $s=1,3,\ldots, 2n -1$, $M(M+1)$ currents with spin $s=2,4,\ldots , 2n$, and $M(M-1)/2$ spin $2n+1$ currents for bosonic generators. Moreover, it has $M^2$ currents with spin $s=3/2,5/2,\ldots , 2n +1/2$ for fermionic generators.
With $M=1$, the spectrum matches with the one in \cite{Candu:2013uya}.

The $M(M-1)$ spin one currents generate two $so(M)$ affine algebras and one of spin two currents is the energy-momentum tensor. 
In appendix \ref{sec:candks}, we obtain the levels of two $so(M)$ affine algebras as
\begin{align}
\ell_1 = t (2 n +1) + 2 n (M-2)\, , \quad
\ell_2   = -  2 t n  - 2 n (M-2) \, ,  \label{l1l2ssoI}
\end{align}
and the total central charge as
\begin{align}
c = \frac{t M (M-1)}{2 (t + M -2 )} - 3 t M n (2 n +1)  + \frac{M (M+1)}{2}  6 n^2 - \frac{M (M-1)}{2} 18 n^2 \, . \label{cssoI}
\end{align}
Here $t$ is the level of  $osp(M (2 n+1)| 2 M n)$ affine algebra.

\subsubsection*{\underline{Type $osp(M (2 n -1)| 2M n)$}}

We next consider the case with $\mathfrak{g} = osp(M (2 n-1)| 2 M n)$, where the superalgebra is 
decomposed as  in \eqref{tsalgebra2}.
We principally embed $osp(1|2)$ in $osp(2 n-1 |2n)$ generated by $\mathbbm{1}_M \otimes B^\sigma_{2n-1|2n}$.
With the $sl(2) \subset osp(1|2)$, we decompose $osp(M (2 n-1)| 2 M n)$ as
\begin{align}
\begin{aligned}
osp(M(2 n-1)| 2 M n ) 
\simeq \, &  M (M-1)  \left(\bigoplus_{s=0}^{n-1}  \rho_{2s} \right)   \oplus M^2 \left(\bigoplus_{s=1}^{2n-1} \rho_{s-1/2} \right)  \\
&\oplus M ( M+1) \left(\bigoplus_{s=1}^{n-1} \rho_{2s-1} \right)    \oplus \frac{M(M+1)}{2} \rho_{2n-1}\, .
\end{aligned}
\end{align}
After the Hamiltonian reduction, the algebra includes $M(M-1)$ currents with spin $s=1,3,\ldots, 2n -1$, $M(M+1)$ currents with spin $s=2,4,\ldots , 2n-2$, and $M(M+1)/2$ spin $2n$ currents for bosonic generators.
Moreover, it has $M^2$ currents with spin $s=3/2,5/2,\ldots , 2n - 1/2$ for fermionic generators.
With $M=1$, the spin content reduces to the one in \cite{Candu:2013uya}.

The $M(M-1)$ spin one currents generate two $so(M)$ affine algebras and one of spin two currents is the energy-momentum tensor. 
In appendix \ref{sec:candks}, we calculate the levels of two $so(M)$ affine algebras as
\begin{align}
\begin{aligned}
&\ell_1  =  4 t n + (M+2) (2n - 1/2) + (M-2) (-1/2) \, , \\ 
&\ell_2 =- 2 t (2 n -1) + (M+2) ( - 2n + 3/2) + (M-2 ) (-1/2) \, , \label{l1l2ssoII}
\end{aligned}
\end{align}
and the total central charge as
\begin{align}
\begin{aligned}
c =& \frac{t M (M+1)}{2 (k + M/2 + 1)} - 6 t M n (2 n - 1)  \\
&- \frac{M (M+1)}{2}  \left(18 (n-1) n + \frac{5}{2} \right) + \frac{M (M-1)}{2} \left(6 (n-1) n-\frac{1}{2} \right) \, ,
\end{aligned}   \label{cssoII}
\end{align}
where $t$ is the level of $osp(M (2 n-1)| 2 M n)$ affine algebra.

\subsubsection*{\underline{Type $osp(4 m n | 2m (2 n +1))$}}

We then study the case with $\mathfrak{g} = osp(4 m n| 2 m (2 n +1))$ with the decomposition \eqref{tsalgebra1}.
We identify the supergravity sector as the $osp(1|2)$ principally embedded in $osp(2 n+1 |2n)$ generated by 
$\mathbbm{1}_{2m} \otimes B^\sigma_{2n+1|2n}$.
We decompose $osp(4 m n| 2 m (2 n +1))$ with the $sl(2) \subset osp(1|2)$ as
\begin{align}
\begin{aligned}
osp(4 m n| 2 m (2 n +1))
\simeq \, & 2 (2 m^2 + m) \left(\bigoplus_{s=0}^{n-1}  \rho_{2s} \right)   \oplus 4 m^2  \left(\bigoplus_{s=1}^{2n} \rho_{s-1/2} \right)  \\
&\oplus 2 (2 m^2 - m)\left(\bigoplus_{s=1}^{n} \rho_{2s-1} \right)    \oplus  (2 m^2 + m) \rho_{2n}\, .
\end{aligned}
\end{align}
The W-algebra after the Hamiltonian reduction has $2 (2 m^2 + m) $ currents with spin $s=1,3,\ldots, 2n -1$, $2 (2 m^2 - m) $ currents with spin $s=2,4,\ldots , 2n$, and $(2 m^2 + m) $ spin $2n+1$ currents for bosonic generators. Moreover, it has $4 m^2 $ currents with spin $s=3/2,5/2,\ldots , 2n +1/2$ for fermionic generators.

The $2 (2 m^2 + m) $ spin one currents generate two $sp(2m)$ affine algebras and one of spin two currents is the energy-momentum tensor. 
In appendix \ref{sec:candks}, we obtain the levels of two $sp(2m)$ affine algebras as
\begin{align}
\ell_1 = t (2 n +1) + 2n (m +1) \, , \quad
\ell_2 = - 2 t n - 2 n (m+1)  \, ,\label{l1l2sspI}
\end{align}
and the total central charge as
\begin{align}
c = \frac{t m (2 m+1)}{t + m + 1} - 12 t m n (2 n+1) + (2 m^2 - m) 6 n^2 - (2 m^2 + m) 18 n^2 \, .   \label{csspI}
\end{align}
Here $t$ is the level of $osp(4 m n| 2 m (2 n +1))$ affine algebra.

\subsubsection*{\underline{Type $osp(4 m n | 2m (2 n -1))$}}

Finally we examine the case with $\mathfrak{g} = osp(4 m n | 2m (2 n -1))$ and with the decomposition \eqref{tsalgebra3}.
As the supergravity sector, we use the $osp(1|2)$ principally embed  in $osp(2 n-1 |2n)$ generated by $\mathbbm{1}_{2m} \otimes B^\sigma_{2n-1|2n}$.
 We decompose $osp(4 m n | 2m (2 n -1))$ with the $sl(2) \subset osp(1|2)$ as
\begin{align}
\begin{aligned}
osp(4 m n | 2m (2 n -1))
\simeq \, &  2 (2 m^2 + m)\left(\bigoplus_{s=0}^{n-1}  \rho_{2s} \right)   \oplus 4 m^2 \left(\bigoplus_{s=1}^{2n-1} \rho_{s-1/2} \right)  \\
&\oplus 2 (2 m^2 - m) \left(\bigoplus_{s=1}^{n-1} \rho_{2s-1} \right)    \oplus  (2 m^2 - m) \rho_{2n-1}\, .
\end{aligned}
\end{align}
After the Hamiltonian reduction, the algebra includes $  2 (2 m^2 + m)$ currents with spin $s=1,3,\ldots, 2n -1$, $  2 (2 m^2 - m)$ currents with spin $s=2,4,\ldots , 2n-2$, and $ 2 m^2 + m$ spin $2n$ currents for bosonic generators. It also has $4 m^2$ currents with spin $s=3/2,5/2,\ldots , 2n - 1/2$ for fermionic generators.

The $2 (2 m^2 + m)$ spin one currents generate two $sp(2m)$ affine algebras and one of spin two currents is the energy-momentum tensor. 
In appendix \ref{sec:candks}, we find the levels of two $sp(2m)$ affine algebras as
\begin{align}
\begin{aligned}
\ell_1 &= t/2 \cdot 2n + (m-1) (2n -1/2) + (m+1) ( - 1/2) \, ,\\
\ell_2 &= - t/2 \cdot (2 n -1) + (m-1) (-2 n +3/2) + (m+1) ( -1/2) \, ,
\end{aligned} \label{l1l2sspII}
\end{align}
and the total central charge as
\begin{align}
\begin{aligned}
c =& \frac{t m (2 m-1)}{t + 2 m  -2} - 6 t m n (2 n-1)  \\
&- (2 m^2 - m) \left(18 (n-1) n + \frac{5}{2} \right) + (2 m^2 + m)\left(6 (n-1) n-\frac{1}{2} \right) \, , 
\end{aligned}   \label{csspII}
\end{align}
where $t$ is the level of $osp(4 m n | 2m (2 n -1))$ affine algebra.

\subsection{OPEs among generators}
\label{sec:superOPEs}

In this subsection, we compute OPEs among generators of the $\mathcal{N}=1$ rectangular W-algebras only with generators of spins $s=1,3/2,2$.%
\footnote{As discussed at the end of subsection \ref{sec:OPEs}, these algebras could be obtained from mother rectangular W-algebras with integer $\lambda$ introduced in \eqref{sB} and quotienting ideals. Similarly, we expect that coset type truncations exist in these cases as well.}
There are two types of them; one is type $osp(M(2 n-1)| 2 M n)$ with $n=1$, which has $so(M)$ symmetry. The other is type $osp(4 m n | 2 m (2 n -1))$ with $n=1$, which has $sp(2m)$ symmetry.
We first study the $\mathcal{N}=1$ W-algebra with $so(M)$ symmetry, and then briefly discuss the one with $sp(2m)$ symmetry. Since expressions become quite complicated compared with the bosonic cases, we only provide the outlines of what we have done.

\subsubsection*{\underline{Type $osp(M (2 n -1)| 2M n)$ with $n=1$}}

We start from the $\mathcal{N}=1$ W-algebra with $so(M)$ symmetry  obtained from the Hamiltonian reduction of $osp(M|2M)$. 
As in the bosonic case, we introduce the $sl(M)$ generators $t^A_M$ expressed by $M \times M$ matrices and decompose $A = (a,\alpha)$ with $a=1,2,\ldots , M(M-1)/2$ and $\alpha = M(M-1)/2 +1 , \ldots , M^2 -1$ such that $t^a_M$ and $t^\alpha_M$ are the anti-symmetric and traceless symmetric matrices, respectively. 
The invariant tensors are defined as in \eqref{tAB} and the normalization of generators is set as $g^{AB}=\delta^{AB}$.
The W-algebra includes two $so(M)$ currents $J^a,K^a$. 
With $\kappa^{ab} = \frac12 \text{tr} (t^a_M t^b_M)$, the OPEs of $J^a,K^a$ can be written as
\begin{align}
J^a (z) J^b (0) \sim \frac{\ell_1 \kappa^{ab} }{z^2} + \frac{i f^{ab}_{~~c} J^c(0)}{z} \, , \quad
K^a (z) K^b (0) \sim \frac{\ell_2 \kappa^{ab}   }{z^2} + \frac{i f^{ab}_{~~c} K^c(0)}{z} \, .
\label{JKOPEs}
\end{align}
The W-algebra also includes the energy-momentum tensor $T$ satisfying
\begin{align}
T (z) T(0) \sim \frac{c/2}{z^4} + \frac{2 T(0)}{z^2} + \frac{T' (0)}{z} \, .
\end{align}
We do not specify the relations among the levels $\ell_1,\ell_2$ and the central charge $c$ for a moment.
The OPEs between the energy-momentum tensor and the $so(M)$ currents are
\begin{align}
T (z) J ^a (0) \sim \frac{J^a (0)}{z^2} + \frac{{ J^a }' (0)}{z} \, , \quad
T (z) K^a (0) \sim \frac{K^a (0)}{z^2} + \frac{{K^a}' (0)}{z} \, .
\end{align}
The algebra includes fermionic generators $G^A$ with spin $3/2$. There are three types, such as, $G= G^0$ in the singlet, $G^a$ in the adjoint, and $G^\alpha$ in the symmetric representations of $so(M)$. For spin two generators, we have the energy-momentum tensor $T$ and charged spin two currents $Q^\alpha$ transforming as the symmetric representation under the $so(M)$ action.  We choose the basis of generators to satisfy
\begin{align}
T(z) G^A (0) \sim \frac{\frac{3}{2} G^A (0)}{z^2} + \frac{{G^A}' (0)}{z} \, , \quad
T(z) Q^\alpha (0) \sim \frac{2 Q^\alpha (0)}{z^2} + \frac{{ Q^\alpha } ' (0)}{z} 
\end{align}
such that the currents are primary w.r.t.~the Virasoro algebra.

We obtain other OPEs among generators by requiring their associativity. 
For similar analysis without matrix extensions but with supersymmetry, see \cite{Candu:2012tr,Candu:2013uya,Beccaria:2014jra}.
The operator products would produce composite operators of spin up to three, and we need to list all independent ones.
In order to avoid making expressions complicated, we use  abbreviated notation for composite operators as
\begin{align}
[A_1 \cdots A_p] \equiv (A_1 \cdots A_p) + \cdots \, ,
\end{align}
where $\dots$ denote terms needed to satisfy the condition primary w.r.t.~the Virasoro algebra.
Here we use the normal ordering prescription given in \eqref{normalorder} 
to make the products of operators well defined.
The definitions of composite operators are not unique in general and this fact will be utilized to make the form of OPEs simpler below.
For spin one and 3/2, there are no primary operators other than the fundamental ones, $J^a$, $K^a$, and $G^A$.  We have primary operators
\begin{align}
[J^{(a} J^{b)}] \, , \quad [K^{(a} K^{b)}] \, , \quad [J^{a} K^{b}] \, , \quad [J^a G^A] \, , \quad [K^a G^A] 
\label{comspin252}
\end{align}
for spin two and 5/2 along with  $Q^a$.
We also find the composite primaries
\begin{align}
\begin{aligned}
&[J^{(a} J^{b} J^{c)}] \, , \quad [K^a J^{(b } J^{c)}] \, , \quad  [K^a J^{(b } J^{c)}] \, , \quad [K^{(a} K^{b} K^{c)}] \, ,  \\ & [J^{[a} { J^{b]} } '] \, , \quad [K^{[a } {K^{b]}}'] \, ,  \quad  [J^{[a} { K^{b]} } ']   \, , \quad 
[J^a Q^\alpha] \, , \quad [K^a Q^\alpha] \, ,\\ & [G G] \, , \quad [G G^a]  \, , \quad [G G^\alpha] \, , \quad [G^{[a } G^{b]}] \, , \quad [G^{a} G^\alpha] \, , \quad [G^{[\alpha} G^{\beta]}]
\end{aligned}
\label{comspin3}
\end{align}
for spin three.

In the following, we consider the specific examples with $M=3,4,5$ except for the OPEs of $Q^a \times Q^b$. Here we have used the symbolic form to express operator products.
The operator product of spin $s_1$ and $s_2$ operators produces (composite) generators of spin up to $s_1 +s_2 -1$, so it is convenient to start from the cases with smaller $s_1 +s_2$.
The smallest case is with $s_1 = s_2 =1$, but the OPEs were already given in \eqref{JKOPEs}.
The next simplest case is with $s_1=1$ and $s_2=3/2$. Denoting $I^P = (J^a,K^a)$, we examine the associativity of $I^P \times I^Q \times G^A$. Up to the normalizations of $G^A$,  the associativity turns out to fix the OPEs of $I^P \times G^A$ uniquely. Firstly, we set
\begin{align}
J^a (z) G(0) \sim \frac{G^a (0)}{z} \, , \quad K^a (z) G(0) \sim  - \frac{G^a (0)}{z} \, ,
\label{JGOPEs}
\end{align}
where the OPEs determine the normalization of $G^a$ relative to $G$. The OPEs of $I^P \times G^a$ are given by
\begin{align}
\begin{aligned}
&J^a (z) G^b (0) \sim \frac{\frac{1}{M} \delta^{ab} G(0) + \frac{i}{2} f^{ab}_{~~c} G^c (0) + \frac{1}{2} d^{ab}_{~~\alpha} G^\alpha (0)}{z} \, , \\
&K^a (z) G^b (0) \sim \frac{- \frac{1}{M} \delta^{ab} G(0) + \frac{i}{2} f^{ab}_{~~c} G^c (0) - \frac{1}{2} d^{ab}_{~~\alpha} G^\alpha (0)}{z} \, .
\end{aligned}
\label{JGaOPEs}
\end{align}
Here we have fixed the normalization of $G^\alpha$ relative to $G$ by the OPEs.
With the normalizations, the OPEs of $I^P \times G^\alpha$ are
\begin{align}
\begin{aligned}
&J^a (z) G^\alpha (0) \sim \frac{ \frac{i}{2} f^{a \alpha}_{~~\, b} G^b (0) + \frac{1}{2} d^{a\alpha}_{~~\beta} G^\beta (0)}{z} \, , \\
&K^a (z) G^\alpha (0) \sim \frac{ \frac{i}{2} f^{a \alpha}_{~~\, b} G^b (0) - \frac{1}{2} d^{a\alpha }_{~~\beta} G^\beta (0)}{z} \, .
\label{JGalphaOPEs}
\end{aligned}
\end{align}
In the above analysis, we used the fact that only the non-trivial invariant tensors are
\begin{align}
f^{abc} \, , \quad f^{a \alpha \beta} \, , \quad  d^{ab \alpha} \, , \quad d^{\alpha \beta \gamma}
\end{align}
among $f^{ABC}$  and $d^{ABC}$.

We next examine the OPEs of $I^P \times Q^\alpha$.
We use the ansatz schematically of the forms
\begin{align}
\begin{aligned}
J^a \times Q^\alpha \sim \, & if ^{a\alpha}_{~~\beta} a_1 Q^\beta + d^{a\alpha}_{~~b} (a_{21} J^b + a_{22} K^b) \\
&+ A^{a\alpha}_{3,bc} [J^{(b} J^{c)}] + A^{a\alpha}_{4,bc} [K^{(b} K^{c)}] + A^{a\alpha}_{5,bc} [J^{b} K^{c}] \, , \\
K^a \times Q^\alpha \sim\, & if ^{a\alpha}_{~~\beta} b_1 Q^\beta + d^{a\alpha}_{~~b} (b_{21} J^b + b_{22} K^b) \\
&+ B^{a\alpha}_{3,bc} [J^{(b} J^{c)}] + B^{a\alpha}_{4,bc} [K^{(b} K^{c)}] + B^{a\alpha}_{5,bc} [J^{b} K^{c}] \, .
\end{aligned}
\end{align}
Here the small letters are constants without indices.
The capital letters are given by linear combinations of invariant tensors as in \eqref{tAB}, \eqref{4tensors}, and \eqref{5tensors} with constant coefficients.
We examine the associativity of $I^P \times I^Q \times Q^\alpha$ and find that there might be several discrete solutions of $a_1$ and $b_1$.
The charge assignments of $Q^\alpha$ w.r.t.~$J^a$ and $K^a$ can be read off from the Hamiltonian reduction of $osp(M|2M)$, and the possible choice is 
$a_1 \neq 0$ and $b_1 = 0$ or $a_1 = 0$ and $b_1 \neq 0$. We use the former one by breaking the symmetry
under the exchange of $J^a$ and $K^a$. 
Moreover, there are ambiguities by redefining the spin two currents $Q^\alpha$ such that
\begin{align}
Q^\alpha \to z_0 Q^\alpha + d^{\alpha}_{~ab} \left(z_1   [J^{(a}J^{b)} ] + z_2   [K^{(a} K^{b)} ] + z_3 [ J^{(a } K^{b)}] \right) 
\label{Qredef}
\end{align}
as mentioned above.
Using $z_1,z_2,z_3$, we set $a_{21}= a_{22} = b_{22} = 0$ to make the OPEs simpler.
With this setup,  solutions to the constraints from  $I^P \times I^Q \times Q^\alpha$ are given by
\begin{align}
J^a (z) Q^\alpha \sim \frac{i f^{a \alpha}_{~~\, \beta} Q^\beta (0)}{z} \, , \quad K^a Q^\alpha \sim 0 \, . \label{JQOPEs}
\end{align}

In a similar manner, we can determine the OPEs of $G^A \times G^B$ and $G^A \times Q^\alpha$ 
in terms of one parameter $\ell_1$ up to the normalizations of $G$ and $Q^\alpha$ (or $z_0$ in \eqref{Qredef}). Here we have used the associativity of $I^P \times G^A \times G^B$, $I^P \times G^A \times Q^\alpha$, and $G^A \times G^B \times G^B$. In particular, we have found the expressions of $c$ and $\ell_2$ in terms of $\ell_1$ for $M=3,4,5$ as 
\begin{align}
c =-\frac{M \left(M^2+M (2 \ell_1-5)+\ell_1 (3 \ell_1 -11)+4\right)}{2 (M+\ell_1)} \, , \quad \ell_2 = 2 - \frac12 (\ell_1 + M) \, ,
\end{align}
which can be obtained from \eqref{l1l2ssoII} and \eqref{cssoII} as well.
We have also determined the OPEs of $Q^\alpha \times Q^\beta$ in terms of one parameter $\ell_1$ up to the normalizations of $G$ and $Q^\alpha$ for the simplest non-trivial example with $M=3$.

\subsubsection*{\underline{Type $osp(4 m n | 2m (2 n -1))$ with $n=1$}}

We then consider the $\mathcal{N}=1$ W-algebra with $sp(2m)$ symmetry obtained from the Hamiltonian reduction of $osp(4 m |2 m)$. 
The situation is quite similar to the previous case, so we explain it only briefly.
As above, we use  $t^A_{2m}$ as the generators of $sl(2m)$ and express them by $2m \times 2m$ matrices.
We decompose the indices as $A =(a,\alpha)$ with $a= 1 ,2 \ldots , 2m^2 + m$ and $\alpha =2 m^2 + m + 1, \ldots , 4 m^2 -1$ such that \eqref{conditionofsp} are satisfied. The invariant tensors are introduced as in \eqref{tAB} with $M =2m$,
and  indices are raised or lowered by $g^{AB}$ or its inverse.
There are two $sp(2m)$ currents $J^a,K^a$ and their levels are denoted as $\ell_1,\ell_2$, respectively, as in \eqref{JKOPEs} but with $\kappa^{ab} = g^{ab}$. There are fermionic generators $G^A$ with spin $3/2$, which are $G = G^0$ in the singlet, $G^a$ in the adjoint, and $G^\alpha$ in the anti-symmetric representations of $sp(2m)$. For spin two generators, we have the energy-momentum tensor $T$ and charged spin two currents $Q^\alpha$ transforming as the anti-symmetric representation under the $sp(2m)$ action. We choose the basis primary w.r.t.~the Virasoro algebra.

We  determine OPEs by requiring their associativity.
Here we mainly consider the examples with $sp(4)$ and $sp(6)$.
For the OPEs of the form $I^P \times G^A$ with $I^P =( J^a,K^a )$, we find
\begin{align}
\begin{aligned}
J^a (z) G^b(0) \sim  \frac{\frac{1}{2m} g^{ab} G(0) + \frac{i}{2} f^{ab}_{~~c} (0) + \frac{1}{2} d^{ab}_{~~\alpha} G^\alpha (0) }{z} \, , \\
K^a (z) G^b(0) \sim  \frac{- \frac{1}{2m} g^{ab} G(0) + \frac{i}{2} f^{ab}_{~~c} (0) - \frac{1}{2} d^{ab}_{~~\alpha} G^\alpha (0) }{z} 
\end{aligned}
\end{align}
along with \eqref{JGOPEs} and \eqref{JGalphaOPEs}.
The composite operators generated by operator products are the same as before as in \eqref{comspin252} and \eqref{comspin3}. We can apply the same arguments as before for the OPEs of $I^P \times Q^\alpha$, and we arrive at the same expressions as in \eqref{JQOPEs}.
The OPEs of $G^A \times G^B$ and $G^A \times Q^\alpha$ can be determined in terms of one parameter, say, $\ell_1$ up to the overall normalizations of $G$ and $Q^\alpha$. The central charge $c$ and the other level $\ell_2$ can be written in terms of $\ell_1$ as 
\begin{align}
c = -\frac{m \left(4 \ell_1 m+\ell_1 (6 \ell_1+11)+2 m^2+5 m+2\right)}{\ell_1+m} \, , \quad \ell_2 = - 1 -  \frac{1}{2} ( \ell_1 + m) 
\end{align}
with $m=2,3$.
The same expressions can be obtained from \eqref{l1l2sspII} and \eqref{csspII}.
We checked that the OPEs of $Q^\alpha \times Q^\beta$ are uniquely fixed in terms of $\ell_1$ up to the normalizations of $G$ and $Q^\alpha$  for  the  example of $sp(4)$.

\subsection{Comparison with dual coset models}
\label{sec:supermap}

In this subsection, we propose cosets realizing the four types of $\mathcal{N}=1$ rectangular W-algebras as their symmetry algebras as listed in table \ref{fig:proposal} and identify the maps of parameters.%
\footnote{See appendix \ref{sec:alternative} for alternative proposals.}
The cosets can be regarded as the holographic duals of the $\mathcal{N}=1$ matrix extensions of  higher spin supergravities. The dual cosets were proposed in \cite{Eberhardt:2018plx} by generalizing the holographic duality in \cite{Creutzig:2013tja} with $\mathcal{N}=2$ supersymmetry.
However, the analysis was done in large $c$ limit and with generic 't Hooft parameters. 
We are interested in the cases with finite $c$ and integer $\lambda$, and,
in particular, the truncations depend on whether $\lambda$ is even or odd.
For this, we work with the four cases contrary to the two cases analyzed in  \cite{Eberhardt:2018plx}.

\subsubsection*{\underline{Type $osp(M (2 n+1) | 2 M  n)$}}

We first consider the $\mathcal{N}=1$ W-algebra with $so(M)$ symmetry obtained as the Hamiltonian reduction of $osp(M (2 n+1) | 2 M  n)$.
We propose that the algebra is realized as the symmetry of the coset
\begin{align}
\frac{so(N + M)_k \oplus so ( N M ) _1 }{so (N)_{k+M}} \, , \label{dualcosetssoI}
\end{align}
where $so ( N M ) _1 $ can be described by $NM$ Majorana fermions.
The same coset was considered in \cite{Eberhardt:2018plx} and also in \cite{Candu:2013uya} with $M=1$.
The symmetry algebra includes two sets of $so(M)$ currents. One comes from $so(N +M)$ and with level $k$.
The other can be constructed from the free fermions and with level $N$.
The central charge is computed as
\begin{align}
c =\frac{k  (N+M)(N+M-1)}{2 (k+N+M-2)}+\frac{NM }{2}  -\frac{(k+M)N (N-1)  }{2 (k+N+M-2)} \, .
\label{centralssoI}
\end{align}
Compared with \eqref{l1l2ssoI} and \eqref{cssoI}, we find a map
\begin{align}
\ell_1 = k \, , \quad \ell_2 = N  \, , \quad  \lambda \equiv   \frac{N}{k+N+M - 2}= - 2n \label{thooftssoI1}
\end{align}
or
\begin{align}
\ell_1 = N \, , \quad \ell_2 = k  \, , \quad \lambda \equiv \frac{k}{k+M+N -2 }  = - 2n \, . \label{thooftssoI2}
\end{align}
The two choices may be understood as a duality of the coset \eqref{dualcosetssoI} by exchanging two sets of $so(M)$ currents.

\subsubsection*{\underline{Type $osp(M (2 n-1) | 2 M  n)$}}

We then examine the $\mathcal{N}=1$ W-algebra with $so(M)$ symmetry given by the Hamiltonian reduction of $osp(M (2 n-1) | 2 M  n)$.
We propose that the algebra is realized as the symmetry of the coset
\begin{align}
\frac{osp(M |2 N)_k \oplus sp ( 2 N M) _{-1/2}}{sp (2 N )_{k-M/2}} \, , \label{dualcosetssoII}
\end{align}
where $sp ( 2 N M) _{-1/2}$ is described by $2 N M$ symplectic bosons.
For $M=1$, the coset reduces to the one proposed in \cite{Candu:2013uya}.
The symmetry algebra includes affine $so(M)$ with level $- 2 k$ from $osp(M|2N)_k$  and 
 affine $so(M)$ with level $-2 N$ from the symplectic bosons.
The central charge is
\begin{align}
c = \frac{k ( M (M -1)/2 +  N(2 N +1) - 2 N M)}{k + N - M/2 + 1} - NM - \frac{(k - M/2) N (2 N +1)}{k -M/2+ N + 1} \, . \label{centralssoII}
\end{align}
The comparison with  \eqref{l1l2ssoII} and \eqref{cssoII} 
leads to a map
\begin{align}
\ell_1 = - 2 k \, , \quad \ell_2 =  - 2 N \, , \quad  \lambda \equiv \frac{N+1}{k+ N-M/2+1} = - 2n +1 
\label{smap1}
\end{align}
or
\begin{align}
\ell_1 =  - 2 N\, , \quad \ell_2 = - 2 k  \, , \quad  \lambda \equiv \frac{k+1}{k+ N-M/2+1} = - 2n +1 \, . 
\label{smap15}
\end{align}
The existence of two choices implies a duality of the coset \eqref{dualcosetssoII}.

\subsubsection*{\underline{Type $osp(4 m n | 2 m (2 n + 1))$}}

One of the $\mathcal{N}=1$ algebras with $sp(2m)$ symmetry can be constructed by the Hamiltonian reduction of $osp(4 m n | 2 m (2 n + 1))$.
We propose that the algebra can be identified as the symmetry  of the coset
\begin{align}
\frac{osp(N | 2 m)_k \oplus sp (  2 N m ) _{-1/2}}{so (N)_{k - 2m}} \, . \label{dualcosetsspI}
\end{align}
The symmetry algebra includes two affine $sp(2m)$ with level $- k /2$ and  level $- N /2$ as subalgebras.
The central charge is
\begin{align}
\begin{aligned}
c = \frac{k \left ( N (N-1) /2 + m (2 m+1)  -2 m N\right)}{k+ N -2 m -2}- N m  -\frac{(k-2 m) N (N-1) }{2 (k+N-2 m-2)} \, .
\label{centralsspI}
\end{aligned}
\end{align}
Compared with \eqref{l1l2sspI} and \eqref{csspI}, we find a map
\begin{align}
\ell_1 = - k /2 \, , \quad \ell_2 =  -  N/2  \, , \quad  \lambda \equiv \frac{N}{k + N -2 m- 2} = - 2n \label{thooftsspI1}
\end{align}
or
\begin{align}
\ell_1 = -  N /2 \, , \quad \ell_2 = - k /2  \, , \quad \lambda \equiv \frac{k}{k +  N-2 m- 2} = - 2n \, .  \label{thooftsspI2}
\end{align}
The two choices imply a duality of the coset \eqref{dualcosetsspI} as above.

\subsubsection*{\underline{Type $osp(4 m n | 2 m (2 n -1))$}}

The other $\mathcal{N}=1$ algebra with $sp(2m)$ symmetry is constructed by the Hamiltonian reduction of $osp(4 m n | 2 m (2 n - 1))$.
Our proposal is that the algebra is realized as the symmetry of the coset
\begin{align}
\frac{sp(2 N + 2m)_k \oplus so ( 4 N m) _1 }{sp (2 N )_{k + m}} \, . \label{dualcosetsspII}
\end{align}
The same coset was considered in \cite{Eberhardt:2018plx}.
The symmetry algebra includes two affine $sp(2 m)$ with level $k$ and  level $N$ as subalgebras.
The central charge is
\begin{align}
c = \frac{k ( (N + m) ( 2(N + m) + 1 )}{k + N + m + 1} + 2 N m - \frac{(k + m ) (2N^2 + N)}{k + m + N +1} \, . \label{centralsspII}
\end{align}
We compare them with \eqref{l1l2sspII}, \eqref{csspII} and find that the correspondence happens at
\begin{align}
\ell_1 = k \, , \quad \ell_2 =  N \, , \quad  \lambda \equiv \frac{N+1}{k+m+N+1} = - 2 n + 1
\label{smap2}
\end{align}
or
\begin{align}
\ell_1 = N \, , \quad \ell_2 = k  \, , \quad \lambda \equiv \frac{k+1}{k+m+N+1} = - 2n + 1\, . 
\label{smap25}
\end{align}
We again expect a duality of the coset \eqref{dualcosetsspII}

\subsection{Symmetry of dual coset models}
\label{sec:superdualcosets}

In subsection \ref{sec:superOPEs}, we obtained  OPEs for the $\mathcal{N}=1$ W-algebras with generators up to spin two.
In this subsection, we explicitly construct generators of the W-algebras in terms of dual coset models and reproduce the OPEs among generators.

\subsubsection*{\underline{Type $osp(M (2 n-1) | M 2 n)$ with $n=1$}}

We first examine the $\mathcal{N}=1$ W-algebra with $so(M)$ symmetry obtained as the Hamiltonian reduction of  $osp(M  |2 M )$. We have proposed that the algebra can be realized as the symmetry of the coset \eqref{dualcosetssoII}.
In order to write down the symmetry generators of the coset \eqref{dualcosetssoII}, we introduce the generators of $osp(M| 2 N)$ as in \eqref{ospgenerators}, \eqref{ospgenerators2} and the invariant tensors as in \eqref{tPQ0}, \eqref{tPQ}.
The osp$(M | 2 N)$ currents consist of bosonic ones $J^a,J^p$ and fermionic ones $\tilde J^i$ and the non-trivial OPEs are given in \eqref{ospOPE}.
We use the description of symplectic bosons $\varphi^i$ for $sp(2 N M)_{-1/2}$ with 
\begin{align}
\varphi^i (z) \varphi^j (0) \sim \frac{g^{ij}}{z} \, .
\end{align}
Using the symplectic bosons, we can construct $so(M)_{-2 N}$ and $sp(2N)_{- M/2}$ currents as
\begin{align}
J_f^a = \frac{i}{2} f^{a}_{~ij} (\varphi^i \varphi^j) \, , \quad 
J_f^p = \frac{i}{2} f^{p}_{~ij} (\varphi^i \varphi^j) \, .
\end{align}
In particular, the $sp(2N)_{k - M/2}$ factor in the coset \eqref{dualcosetssoII} is given by
\begin{align}
\hat J^p = J^p + J_f^p \, .
\end{align}

In terms of these currents, we write down generators of the $\mathcal{N}=1$ W-algebra with  $so(M)$ symmetry up to spin two. For spin one generators, we have $J^a$ and $K^a = J^a_f$ with levels $-2 k$ and $-2 N$, respectively.
For spin 3/2 generators, we use products of $\tilde J^j$ and $\varphi^i$ such as to be regular w.r.t.~$\hat J^p$.
Using suitable invariant tensors, we have
\begin{align}
G = g_{ij} (\tilde J^i \varphi^j) \, , \quad G^a = i f^{a}_{~ij} (\tilde J^i \varphi^j) \, , \quad G^\alpha = - d^{\alpha}_{~ij} (\tilde J^i \varphi^j) \, ,
\end{align}
where the overall factors are chosen in order to match with the OPEs \eqref{JGOPEs}, \eqref{JGaOPEs}, and \eqref{JGalphaOPEs}.
There are spin two generators, such as, the energy-momentum tensor $T$ and  charged spin two currents $Q^\alpha$. The energy-momentum tensor can be obtained by the standard coset construction \cite{Goddard:1984vk}.
For the charged spin two generators, we find that
\begin{align}
Q^\alpha =  d^{\alpha }_{~ij} (\tilde J^i \tilde J^j) + \frac{2 N   }{4 k - M + 4 } d^{\alpha}_{~ab} (J^{a} J^{b})  
\end{align}
are Virasoro primaries and satisfy the OPEs \eqref{JQOPEs}.
They are actually the same as those for the bosonic case in \eqref{tQso}, and this is because we use the definition of spin two currents such as to satisfy the same OPEs.
Setting
\begin{align}
\ell_1 = - 2 k \, , \quad \ell_2 = - 2 N \, , \, \quad k = - 2 N + M/2 - 2
\end{align}
as in \eqref{smap1} with $n=1$, we have checked for several examples that other OPEs among generators are reproduced up to null vectors. 

\subsubsection*{\underline{Type $osp(4 m n | 2 m (2 n -1))$ with $n=1$}}

We then examine the $\mathcal{N}=1$ W-algebra with $sp(2m)$ symmetry obtained from   $osp(4 m | 2 m )$, which is supposed to be realized as the symmetry of the coset \eqref{dualcosetsspII}.
In order to describe the currents in the coset \eqref{dualcosetsspII},
we use  the generators of $sp( 2m + 2 N)$ as in \eqref{spgenerators}, \eqref{spgenerators2} and the invariant tensors as in \eqref{tPQ0}, \eqref{tPQ}.
With the notation, the OPEs of $sp(2m+2N)$ currents and the free fermions $\psi^i$ from $so(4 m N)_1$ are given by \eqref{spOPE} and 
\begin{align}
\psi^i (z) \psi^j (0) \sim \frac{g^{ij}}{z} \, .
\end{align}
We can construct $sp(2m)_{N}$ and $sp(2N)_{m}$ currents from the free fermions as
\begin{align}
J_f^{a} = - \frac{i}{2} f^{a}_{~ ij} (\psi^i \psi^j) \, , \quad
J_f^p = - \frac{i}{2} f^{p}_{~ij} (\psi^i \psi^j) \, .
\end{align}
The $sp(2N)_{k+m}$ currents in the coset \eqref{dualcosetsspII} are given by
\begin{align}
\hat J^p = J^p + J^p_f \, .
\end{align}

Next we construct generators of the $\mathcal{N}=1$ W-algebra up to spin two in the coset language.
Spin one generators are $J^a$ and $K^a = J_f^a$. Spin 3/2 generators are constructed as
\begin{align}
G = g_{ij} (\tilde J^i \psi^j) \, , \quad G^a = - i f^{a}_{~ij} (\tilde J^i \psi^j) \, , \quad
G^\alpha = d^{\alpha}_{~ij} (\tilde J^i \psi^j) 
\end{align}
as in the case with $so(M)$. As a spin two current, the energy-momentum tensor $T$ can be constructed in the standard way \cite{Goddard:1984vk}. The other spin two currents are given by
\begin{align}
Q^\alpha = d^{\alpha}_{~ij} (\tilde J^i \tilde J^j) - \frac{N}{2 k + m + 2 } d^{\alpha}_{~ab} ( J^a J^b) \, ,
\end{align}
which are primary w.r.t the Virasoro algebra and satisfy the OPEs \eqref{JQOPEs}.
They are the same as \eqref{tQsp} in the bosonic case because of our definition of spin two currents. 
Setting
\begin{align}
\ell_1 = k \, , \quad \ell_2 = N \, , \quad k = - 2 N - m - 2 
\end{align}
as in \eqref{smap2} with $n=1$,
we have checked  for several examples that other OPEs among generators are reproduced up to null vectors.

\subsection*{Acknowledgements}

YH thanks the organizers of ESI Programme and Workshop ``Higher spins and holography'' at the Erwin Schr{\"o}dinger Institute in Vienna, where a part of this work was done.
The work of TC is supported by NSERC grant number RES0019997.
The work of YH is supported by JSPS KAKENHI Grant Number 16H02182 and 19H01896.
The work of TU is supported by the Grant-in-Aid for JSPS Research Fellow, No.19J11212.

\appendix

\section{Computations on central charges and levels}
\label{sec:candk}

In this appendix, we write down some details of computations on the central charges of rectangular W-algebras and the levels of $so(M)$ or $sp(2m)$ affine subalgebras.
Let us consider a vertex algebra $W_t (\mathfrak{g},\hat x, \hat f)$.
Here $\mathfrak{g}$ denotes a Lie (super)algebra with a suitable norm $(.|.)$.
The label $t$ is the level of the universal affine vertex algebra $V_t (\mathfrak{g})$, which is used to construct  $W_t (\mathfrak{g},\hat x,\hat f)$.
Furthermore, $\hat x, \hat f$ are even elements of $\mathfrak{g}$, and
we require that they form a sl$(2)$ algebra \eqref{sl2comm}
with an additional element $\hat e$.
We decompose $\mathfrak{g}$ by the eigenvalue of adjoint action ad$\, \hat x$ as
\begin{align}
	\mathfrak{g} = \oplus_{j \in \frac12 \mathbb{Z}} \mathfrak{g}_j \, .
\end{align}
We choose a basis $\{ u_\alpha \}_{\alpha \in S_j}$ for $\mathfrak{g}_j$ and denote $S_+ = \prod_{j > 0} S_j$. The formula of central charge for the vertex algebra $W_t (\mathfrak{g},\hat x,\hat f)$ may be found in
(2.3) of \cite{Kac:2003jh} as
\begin{align}
	c =& \frac{t \, \text{sdim} \, \mathfrak{g}}{t + h^\vee} - 12 t (\hat x|\hat x)  - \sum_{\alpha \in S_+} (-1)^{p (\alpha)} (12 m^2_\alpha  - 12 m_\alpha + 2) - \frac12 \, \text{sdim} \,  \mathfrak{g}_{1/2} \, .
	\label{KW}
\end{align} 
Here $h^\vee$ represents the dual Coxeter number of $\mathfrak{g}$ and $p(\alpha)$ denotes the parity of $u_\alpha$. Moreover, $(m_\alpha, 1 - m_\alpha)$ are the conformal dimensions of ghost system and  we set  $m_\alpha = 1 - j$. There is no universal formula for the levels of affine subalgebras, but they can be evaluated  for each  example as below.

\subsection{Rectangular W-algebras}
\label{sec:candkb}

As explained in subsection \ref{sec:basic}, there are four types of bosonic rectangular W-algebras obtained from the Hamiltonian reductions of the gauge algebras listed in table \ref{fig:hsa}.
Here we compute the central charges of the algebras by applying the formula \eqref{KW} and moreover obtain the levels of $so(M)$ or $sp(2m)$ affine subalgebras.

\subsubsection*{\underline{Type $so(M(2 n +1))$}}

We first consider the W-algebra with $so(M)$ symmetry obtained as the Hamiltonian reduction of $so(M(2 n +1))$.
We decompose $so(M(2 n +1))$ as in \eqref{talgebra0} or \eqref{decsoI} and use the $sl(2)$ principally embedded in  $\mathbbm{1}_M \otimes so(2 n+ 1)$.

The central charge of the algebra can be computed by applying the formula \eqref{KW}.
The dimension of $so(M (2 n +1))$ is $M (2 n+1) (M (2 n+1)-1)/2 $ and the dual Coxeter number is $M (2 n + 1)  -2$.
We use the convention such that $(\hat y|\hat z) = \frac12 \text{tr} (\hat y \hat z)$ to match with the common one. Using this, we find
\begin{align}
(\hat x|\hat x) =  \frac12 \cdot M \cdot \frac{1}{12}  (2 n+1) ((2 n +1)^2 -1)  = \frac{1}{6} Mn (n+1) (2 n+1) \, . 
\end{align}
We them compute the number of elements in $S_j$.
With the principal embedding of $sl(2)$, the elements of $so( 2 n + 1)$ can be decomposed as
\begin{align}
so(2 n + 1) =  \bigoplus_{s=1}^n \rho_{2s-1} \, . \label{so2n+1dec}
\end{align}
Moreover, the number of elements of symmetric matrix $A_M^s$ is $C_M^s$ in \eqref{CM}.
From this, we can read off the number of ghost system from the sector as $C_M^s(2 n+ 1 - j)/2$ for $j=1,3,\ldots , 2 n - 1$ and $C_M^s(2 n  - j )/2$ for $j=2,4,\ldots,2 n -2$. The contribution to the central charge is
\begin{align}
	\begin{aligned}
		&- C_M^s \sum_{p = 0}^{n-1} [(n - p) (12 (2 p +1)^2 - 12 (2 p + 1) + 2] \\
		&- C_M^s\sum_{p = 0}^{n-2} [(n - 1-  p) (12 (2 p +2)^2 - 12 (2 p + 2) + 2] \\
		&=  C_M^s (6 n^2 - 8 n^4 ) \, .
	\end{aligned}
\end{align}
In a similar way, we decompose the symmetric representation of $so(2 n +1)$ as
\begin{align}
 (\text{sym})_{2 n +1} = \bigoplus_{s=1}^n \rho_{2s} \, . \label{sym2n+1dec}
\end{align}
The number of elements of anti-symmetric matrix $A_M^a$  is $C_M^a$ in \eqref{CM}.
The number of ghost system from the sector is $C_M^a (2 n +1 - j)/2$ for $j=1,3,\cdots , 2 n -1$ and $C_M^a (2 n - j +2)/2$ for $j=2,4,\cdots, 2 n$.  The contribution to the central charge is
\begin{align}
	\begin{aligned}
		&-C_M^a \sum_{p = 0}^{n -1} [(n - p) (12 (2 p +1)^2 - 12 (2 p + 1) + 2] \\
		&- C_M^a\sum_{p = 0}^{n - 1} [(n -  p) (12 (2 p +2)^2 - 12 (2 p +2 ) + 2] \\
		&=-C_M^a  2 n (n+1) (4 n (n+1)-1) \, .
	\end{aligned}
\end{align}
The central charge is thus obtained  as \eqref{csoI}.

We next compute the level of $so(M)$ affine algebra.
From the diagonal sector of $so(M) \otimes \mathbbm{1}_{2n +1}$, we have $so(M)$ currents with the level $t (2 n +1) $.
Furthermore, there are ghosts, which transform in the symmetric and adjoint representations of $so(M)$, and
we can construct $so(M)$ currents from the ghost system.
The level of the  $so(M)$ currents is $M+2$ (or $M-2$) for a set of ghosts in the symmetric (or adjoint) representation of $so(M)$, thus the total $so(M)$ currents receive the shifts of level from a set by $M+2$  (or $M-2$).
The number of the sets is $ 2 \sum_{s=1}^n (2 s -1)  = 2 n^2$ (or $2 \sum_{s=1}^n (2 s) =  2 n (n+1)$) for the symmetric (or adjoint) representation of $so(M)$.
From this, the level of $so(M)$ currents is obtained as \eqref{lsoI}.

\subsubsection*{\underline{Type $sp(2 M n )$}}

We then examine the W-algebra with $so(M)$ symmetry obtained as the Hamiltonian reduction of $sp (2 M n)$.
We decompose $sp (2 M n)$ as in \eqref{talgebra2} or \eqref{decsoII} and use the $sl(2)$ principally embedded in  $\mathbbm{1}_M \otimes sp(2 n)$.

We compute the central charge from \eqref{KW}.
The dimension of $sp(2 M n)$ is $ M n (2 M n + 1)$ and the dual Coxeter number is $M n  + 1$.
Using the convention $(\hat y|\hat z) = \text{tr} (\hat y \hat z)$, we find
$(\hat x|\hat x) =2 M  n (4 n^2 -1)/12 $. 
We decompose $sp(2 n)$ by the principally embedded $sl(2)$ as
\begin{align}
sp(2 n ) = \bigoplus_{s= 1}^n \rho_{2s-1}  \label{sp2ndec}
\end{align}
and associate the factor $C_{M}^{s}$ in \eqref{CM}. 
Thus, the number of ghost pair is $C_{M}^{s} (2 n +1 - j)/2 $ for $j=1,3,5,\ldots, 2 n -1$ and $C_{M}^{s} (2 n - j)/2 $ for $j=2,4,6,\ldots, 2 n -2$.
The contribution to the central charge from the sector is
\begin{align}
C_M^s (6 n^2 - 8 n^4) \, .
\end{align}
Similarly, we decompose the anti-symmetric representation of $sp(2n)$ by the $sl(2)$ as
\begin{align}
(\text{asym})_{2n} = \bigoplus_{s=1}^{n-1} \rho_{2s } \label{asym2ndec}
\end{align}
 and associate the factor $C_{M}^{a}$ in \eqref{CM}.
The number of ghost pair from the sector is $C_{M}^{a} (2 n - 1 - j)/2 $ for $j=1,3,5,\ldots, 2 n -3$ and $C_{M}^{a} (2 n - j )/2 $ for $j=2,4,6,\ldots, 2 n -2$.
The contribution to the central charge from the sector is 
\begin{align}
-  C_M^a 2 n ( 1 + 3 n - 8 n^2 + 4 n^3) \, .
\end{align}
The total central charge is then evaluated as in \eqref{csoII}.

The level of $so(M)$ affine algebra can be obtained as in the case of \eqref{lsoI}.
The ghosts transform in the symmetric and adjoint representations of $so(M)$, and the numbers of sets of ghosts are $ 2 n^2$ and $2 n (n-1)$, respectively.
From this, the level of $so(M)$ can be found as  \eqref{lsoII}. 
The factor $2$ in the first term of the right hand side comes from the difference of convention for the levels of $so(L)$ and $sp(2K)$.

\subsubsection*{\underline{Type $sp(2 m (2 n +1))$}}

Here we deal with the W-algebra with $sp(2m)$ symmetry from the Hamiltonian reduction of $sp(2 m (2 n+1))$ with the decomposition \eqref{talgebra1} or \eqref{decspI}. We use the principally embedded $sl(2)$ in $\mathbbm{1}_{2m} \otimes so(2n+1)$.
For $sp(2 m (2 n+1))$, the dimension  is $m (2 n +1) (2 m (2 n +1) + 1 )$ and the dual Coxeter number is
$m (2 n +1) + 1$.
The norm is 
$(\hat x|\hat x) = 2 m n (n+1) (2 n + 1 ) /3$. 
We decompose  $so(2n+1)$ by the $sl(2)$ as in \eqref{so2n+1dec} and associate the factor $C_{2m}^{\Omega,s}$ in \eqref{C2m}.
Similarly, we decompose the symmetric representation of $so(2n+1)$ by the $sl(2)$ as in \eqref{sym2n+1dec}
and associate the factor $C_{2m}^{\Omega,a}$ in \eqref{C2m}.
The total central charge  is then evaluated as in \eqref{cspI}.
The ghosts transform in the anti-symmetric and adjoint representations of $sp(2m)$. 
The shifts of level from a set of ghost are $m-1$ and $m+1$, and the numbers of set are $2 n^2$ and $2 n (n + 1)$, respectively.
Therefore, the level of $sp(2m)$ currents is obtained as in \eqref{lspI}.

\subsubsection*{\underline{Type $so(4 m n)$}}

The final example is the W-algebra with $sp(2m)$ symmetry from  the Hamiltonian reduction of $so(4 m n)$ with the decomposition  \eqref{talgebra3} or \eqref{decspII} and with the $sl(2)$ principally embedded in $\mathbbm{1}_{2m} \otimes sp(2n)$.
For $so(4 m n)$, the dimension is $2 m n (4 m n -1)$ and the dual Coxeter number is $4 m n - 2$.
The norm is $(\hat x|\hat x) = 2 m \cdot 2 n (4 n^2 -1)/24$.
The adjoint representation of $sp(2n)$ can be decomposed by the $sl(2)$ as in \eqref{sp2ndec} 
and the factor $C_{2m}^{\Omega,s}$ in \eqref{C2m} is associated.
Similarly, the anti-symmetric representation of $sp(2n)$ is decomposed by the $sl(2)$ as in \eqref{asym2ndec} and the factor $C_{2m}^{\Omega,a}$ in \eqref{C2m} is associated.
The total central charge is then computed as in \eqref{cspII}.
The ghosts transform in the anti-symmetric and adjoint representations of $sp(2m)$, 
and the numbers of sets of ghosts are $2 n^2$ and $2 n (n -1)$, respectively.
Therefore, the level of $sp(2m)$ currents is obtained as in  \eqref{lspII}.

\subsection{$\mathcal{N}=1$ rectangular W-algebras}
\label{sec:candks}

In subsection \ref{sec:superbasics}, we construct four types of $\mathcal{N}=1$ rectangular W-algebras from the Hamiltonian reductions of the superalgebras listed in table \ref{fig:hsa}. 
Here we compute the central charges of the W-algebras and the levels of two $so(M)$ or $sp(2m)$ affine subalgebras.

\subsubsection*{\underline{Type $osp(M (2 n+1) | 2 M  n)$}}

We first examine the $\mathcal{N}=1$ W-algebra with $so(M)$ symmetry obtained as the Hamiltonian reduction of $osp(M (2 n+1) | 2 M  n)$. We express the generators of $osp(M (2 n+1) | 2 M  n)$ by $T_M \otimes U_{2n+1|2n}$, where $T_M$ are $M \times M$ matrices and $U_{2n+1|2n}$ are  $((2 n +1) + 2n) \times ((2 n+1) + 2 n)$ even supermatrices. We assign
\begin{align}
	\Omega_M^{-1} T^\text{tr}_M \Omega_M =  \epsilon T_M \, ,   \quad  \Omega_M^\text{tr} = \Omega_M \, ,  \label{scond1}
\end{align}
and
\begin{align}
	\begin{aligned}
		&\Omega_{2 n +1}^{-1} A_{2 n +1}^\text{tr} \Omega_{2 n +1} = - \epsilon A_{2 n +1} \, ,  
		\quad \Omega_{2n}^{-1} D^\text{tr}_{2n} \Omega_{2n} = - \epsilon D_{2n } \, , \\ &\Omega_{2n}^{-1} B^\text{tr}_{2 n +1 , 2n}  \Omega_{2n+1} = \epsilon C_{2n , 2 n+1}  \, , 	\end{aligned}
\label{scond2}
\end{align}
where
\begin{align}
	U_{2 n+ 1 | 2n } = \begin{pmatrix}
		A_{2 n +1} & B_{2 n +1 , 2n }  \\
		C_{2n , 2 n+1 }  & D_{2n }
	\end{pmatrix} \, , \quad \Omega_{2 n +1} ^\text{tr} =  \Omega_{2 n +1} \, , \quad \Omega_{2n} ^\text{tr} =  - \Omega_{2n} \, .\label{scond3}
\end{align}
With this description, the subalgebras $so(M) \otimes \mathbbm{1}_{2n+1|2n}$ and $\mathbbm{1}_M \otimes osp(2n+1|2n)$ are manifestly realized, where $\mathbbm{1}_{2n+1|2n}$ means \eqref{dec1m2n}.
We adopt the principal embedding of $osp(1|2)$ in the  $\mathbbm{1}_M \otimes osp(2n+1|2n)$ subalgebra. 

We compute the central charge of the algebra by applying the formula \eqref{KW}.
The superdimension of $osp(M (2 n+1) | 2 M  n)$ is
\begin{align}
\begin{aligned}
 \frac{M (2 n +1) (M (2 n+1) - 1)}{2} + M n (2 M n + 1) - M^2 2 n (2 n+ 1) 
&= \frac{ M (M-1)}{2} \, ,
\end{aligned}
\end{align}
and the dual Coxeter number is $h^\vee = M (2 n +1) - 2 M n -2 =  M -2 $.
The norm $(\hat x|\hat x) = \frac12 \text{str} (\hat x \hat x)$ is 
\begin{align}
(\hat x|\hat x)  = \frac{M}{2} \left[ \frac{1}{12} (2 n +1) ((2 n +1)^2 - 1) - \frac{1}{12} 2n (4n^2 -1) \right] = \frac{1}{4} M n (2 n +1) \, .
\end{align}
We count the number of elements in $S_j$ with the choice of $sl(2)$-embedding.
For $\epsilon = +1$ in \eqref{scond1} and \eqref{scond2}, we decompose $osp(2n+1|2n) $ as
\begin{align}
\left( \bigoplus_{s=1}^n  2 \rho_{2s -1}  \right) 
\oplus \left( \bigoplus_{s=1}^{2n} \rho_{s-1/2} \right) \, . \label{ospdec}
\end{align}
Thus, the number of even elements in $S_j$ is $C_M^s (2n +1 -j)$ for $j=1,3,\ldots , 2 n -1$ and $C_M^s (2 n -j)$ for $j=2,4,\cdots , 2 n -2$, where $C_M^s$ is given in \eqref{CM}.
Similarly, the number of odd elements is $C_M^s (2n - j + 1/2)$ for $j=1/2,3/2,\ldots,2n - 1/2$.
For $\epsilon = -1$ in  \eqref{scond1} and \eqref{scond2}, we decompose the elements of $U_{2n+1|2n} $ as
\begin{align}
\left( \bigoplus_{s=0}^{n-1}  2 \rho_{2s} \right) \oplus \rho_{2n} \oplus \left( \bigoplus_{s=1}^{2n}  \rho_{s-1/2} \right) \, . \label{ospdec2}
\end{align}
Thus, the number of even elements in $S_j$ is $C_M^a (2n  -j)$ for $j=1,3,\ldots , 2 n -1$ and $C_M^a (2 n -j + 1)$ for $j=2,4,\cdots , 2 n$, where $C_M^a$ is given in \eqref{CM}. 
Similarly, the number of odd elements is $C_M^a (2n - j + 1/2)$ for $j=1/2,3/2,\ldots,2n - 1/2$.
The total central charge is then computed as in \eqref{cssoI}.

The levels of two sets of $so(M)$ currents can be obtained from the information of ghosts as in the bosonic cases.
A set of $so(M)$ currents come from $so(M) \otimes \mathbbm{1}_{2 n+1}$, and the ghosts from  $so(M(2n+1)) \subset osp(M(2 n+1)|2 Mn)$ give rise to the shift of level by 
$2 (M + 2) n^2 + 2 (M - 2) n (n + 1) $.
Similarly, the other set come from $so(M) \otimes \mathbbm{1}_{2n}$, and the ghosts from $sp(2 M n) \subset osp(M(2n+1)|2Mn)$ contribute to the shift of level by $2 (M+2) n^2 + 2 (M -2) n (n-1) $.
There are bosonic ghost system arising from the off-diagonal blocks of $osp(M(2n+1)|2Mn)$, 
which transform in the vector representations of $so(2n+1)$ and $sp(2n)$ and in the symmetric and adjoint representations of $so(M)$.
A set of bosonic ghosts in the symmetric and adjoint representations of $so(M)$ yield the shifts of level by $-(M+2)$ and $-(M-2)$, respectively.
Thus the shifts of level from the sector are $- (M+2) n(2 n+1) - (M-2) n (2 n +1)$ for the both affine $so(M)$.
There are also $n$ fermionic ghost systems of conformal weight $(1/2,1/2)$ in the symmetric and adjoint representations of $so(M)$, and the contributions to the levels are $(M+2) n + (M -2 ) n $. 
In total, the levels of two sets of $so(M)$ currents are obtained
\begin{align}
\begin{aligned}
\ell_1 =& t (2 n+1) + (M + 2) ( 2 n^2 - n (2 n+1) + n) \\
 &\qquad + (M-2) ( 2 n (n+1) - n (2 n+1) + n) \, ,  \\
\ell_2 =& - t \cdot 2n + (M + 2) ( 2 n^2 - n (2 n+1) + n) \\
& \qquad + (M-2) ( 2 n (n-1) - n (2 n+1) + n) \, ,
\end{aligned}
\end{align}
which lead to \eqref{l1l2ssoI}.

\subsubsection*{\underline{Type  $osp(M (2 n-1) | 2 M  n)$}}

We then move to the $\mathcal{N}=1$ W-algebra with $so(M)$ symmetry as the Hamiltonian reduction of $osp(M (2 n-1) | 2 M  n)$.
We express the generators of $osp(M (2 n-1) | 2 M  n)$ by $T_M \otimes U_{2n-1|2n}$, where  $U_{2 n-1| 2n}$ is given by
\begin{align}
	U_{2 n -1 | 2n} = \begin{pmatrix}
		A_{2n -1 } & B_{2 n -1 , 2n} \\
		C_{2n , 2 n-1} & D_{2n } 
	\end{pmatrix} \, .
\end{align}
We assign the conditions just like for \eqref{scond1}, \eqref{scond2}, and \eqref{scond3}.
With this description, the subalgebras $so(M) \otimes \mathbbm{1}_{2n-1|2n}$ and $\mathbbm{1}_M \otimes osp(2n-1|2n)$ are manifestly realized.
We adopt the principal embedding of $osp(1|2)$ into the  $\mathbbm{1}_M \otimes osp(2n-1|2n)$ subalgebra. 

We compute the central charge from \eqref{KW}.
The superdimension is $\text{sdim } osp(M (2 n-1) | 2 M  n) =  M (M+1) /2$ and the dual Coxeter number is $h^\vee = M/2 + 1$.
The norm is $(\hat x|\hat x) = \text{str} (\hat x \hat x) =  M \cdot 2n (2n -1 ) /4 $.
We examine the number of elements in $S_j$.
For $\epsilon = +1$, we decompose $osp(2n-1|2n)$  as
\begin{align}
\left(  \bigoplus_{s=1}^{n-1} 2 \rho_{2s-1} \right) \oplus \rho_{2 n-1} \oplus \left( \bigoplus_{s=1}^{2n-1} \rho_{s-1/2} \right) \, . \label{ospdec3}
\end{align}
Thus, the number of even elements in $S_j$ is $C_M^s (2n  -j)$ for $j=1,3,\ldots , 2 n -1 $ and $C_M^s (2 n -j -1)$ for $j=2,4,\cdots , 2 n -2$. Similarly, the number of odd elements is $C_M^s (2n - j - 1/2)$ for $j=1/2,3/2,\ldots,2n - 3/2$.
For $\epsilon = -1$, we decompose the elements of $U_{2 n-1|2n } $ as 
\begin{align}
\left( \bigoplus_{s=0}^{n-1}   2 \rho_{2s} \right) \oplus \left(  \bigoplus_{s=1}^{2n-1} \rho_{s-1/2} \right) \, . \label{ospdec4}
\end{align}
Thus, the number of even elements in $S_j$ is $C_M^a (2n  -j -1)$ for $j=1,3,\ldots , 2 n - 3$ and $C_M^a (2 n -j )$ for $j=2,4,\cdots , 2 n -2$. Similarly, the number of odd elements is $C_M^a (2n - j - 1/2)$ for $j=1/2,3/2,\ldots,2n - 3/2$.
The total central charge is given by \eqref{cssoII}.

We can compute the levels of two sets of $so(M)$ currents as above.
We obtain 
\begin{align}
\begin{aligned}
	\ell_1 =&  2 t \cdot 2n + (M + 2) ( 2 n^2 - n (2 n-1) + n - 1/2)  \\
	& \qquad + (M-2) ( 2 n (n-1) - n (2 n-1) + n - 1/2)   \, , \\
	\ell_2 =& - 2 t (2 n-1) + (M + 2) ( 2 (n-1)^2 - n (2 n-1) + n - 1/2)   \\ & \qquad + (M-2) ( 2 n (n-1) - n (2 n-1) + n - 1/2) \, , \\
\end{aligned}
\end{align}
which become \eqref{l1l2ssoII}.

\subsubsection*{\underline{Type  $osp(4 m n | 2m (2 n +1))$}}

Here we consider the $\mathcal{N}=1$ W-algebra with $sp(2m)$ symmetry from the Hamiltonian reduction of $osp(4 m n | 2m (2 n +1))$.
We express the generators of $osp(4 m n | 2 m (2 n +1))$ by $T_{2m} \otimes U_{2 n | 2n +1}$. Here $T_{2m}$ is a $2 m \times 2m$ matrix satisfying 
\begin{align}
	\Omega_T^{-1} T^\text{tr}_{2 m} \Omega_{2 m} =  \epsilon T_{2 m} \, ,   \quad  \Omega_{2 m}^\text{tr} = - \Omega_{2m }  \label{spcond1}
\end{align} 
with $\epsilon = \pm 1$.
We also define $U_{2 n | 2n + 1}$ as
\begin{align}
	\begin{aligned}
		&\Omega_{2n}^{-1} A_{2n }^\text{tr} \Omega_{2n} = - \epsilon A_{2n } \, ,  \quad
		\Omega_{2n+1}^{-1} D^\text{tr}_{2 n +1} \Omega_{2n+1} = - \epsilon D_{2 n +1} \, , \\ &\Omega_{2n+1}^{-1} B^\text{tr}_{2n , 2  n + 1}  \Omega_{2n} = \epsilon C_{2 n +1 , 2n}\label{spcond2}
	\end{aligned}
\end{align}
with 
\begin{align}
	U_{2n | 2 n +1} = \begin{pmatrix}
		A_{2n } & B_{2n , 2 n +1 }  \\
		C_{2 n +1 , 2 n}  & D_{2 n +1 ,2 n +1}
	\end{pmatrix} \, , \quad \Omega_{2n} ^\text{tr} =  - \Omega_{2n} \, , \quad \Omega_{2n+1} ^\text{tr} =  \Omega_{2n+1} \, .\label{spcond3}
\end{align}
With this description, the subalgebras $sp(2m) \otimes \mathbbm{1}_{2n+1|2n}$ and $\mathbbm{1}_{2m} \otimes osp(2n+1|2n)$ are manifestly realized.
We principally embed $osp(1|2)$ into the $\mathbbm{1}_{2m} \otimes osp(2n+1|2n)$ algebra. 
The superdimension is $\text{sdim }  osp(4 m n | 2m (2 n +1)) = m (2 m +1)$ and the dual Coxeter number is $h^\vee =  m + 1$. The norm is $(\hat x|\hat x) = 2 m \cdot 2 n (2 n +1)/4$.
The decompositions of $U_{2n|2n+1}$ with $\epsilon = \pm 1$ by the embedded $sl(2) \subset osp(1|2)$ are the same as in \eqref{ospdec} and \eqref{ospdec2}.
For the number of elements of $S_j$, we need to replace $C_M^{s,a}$ by $C_{2m}^{\Omega,s,a}$ defined in \eqref{C2m}.
The central charge is computed as in \eqref{csspI}, and 
the levels of two sets of $sp(2m)$ currents are \eqref{l1l2sspI}.

\subsubsection*{\underline{Type $osp(4 m n | 2m (2 n - 1))$}}

Finally we deal with the $\mathcal{N}=1$ W-algebra with $sp(2m)$ symmetry from the Hamiltonian reduction of $osp(4 m n | 2m (2 n - 1))$.
We express the generators of $osp(4 m n | 2m (2 n -1))$ by $T_{2m} \otimes U_{2n | 2n -1}$, where  $U_{2n | 2n -1}$ is given by
\begin{align}
	U_{2n | 2n -1} = \begin{pmatrix}
		A_{2n } & B_{2n , 2 n -1} \\
		C_{2 n -1 , 2n} & D_{2 n -1}
	\end{pmatrix}  \, .
\end{align}
We assign the conditions just like for \eqref{spcond1}, \eqref{spcond2}, and \eqref{spcond3}.
With this description, the subalgebras $sp(2m) \otimes \mathbbm{1}_{2n|2n-1}$ and $\mathbbm{1}_{2m} \otimes osp(2n-1|2n)$ are manifestly realized.
We adopt the principal embedding of $osp(1|2)$ into the  $\mathbbm{1}_M \otimes osp(2n-1|2n)$ subalgebra. 
The superdimension is $\text{sdim } osp(4 m n | 2m (2 n - 1)) =m ( 2 m -1)$ and the dual Coxeter number is $h^\vee = 2m  -2$.
The norm is $(\hat x|\hat x) = 2m \cdot 2n (2n -1 ) /8 $.
The decompositions of $U_{2n|2n-1}$ with $\epsilon = \pm$ by the embedded $sl(2) \subset osp(1|2)$ are the same as in \eqref{ospdec3} and \eqref{ospdec4}.
For the number of elements of $S_j$, we need to replace $C_M^{s,a}$ by $C_{2m}^{\Omega,s,a}$.
Thus the central charge is \eqref{csspII} and
the levels of  two sets of $sp(2m)$ currents are \eqref{l1l2sspII}.

\section{Alternative proposals of dual coset models}
\label{sec:alternative}

In this appendix, we provide alternative proposals on the coset models dual to higher spin gravities with restricted matrix extensions as listed in table \ref{fig:proposal2}. Compared with the previous proposals  in table \ref{fig:proposal}, the cosets are simply swapped.
\begin{table}
	\centering
	\begin{align*}
	\begin{array}{ll}
	\text{Type of W-algebra} & \text{Dual coset model} \\ \hline \hline
	\displaystyle so(M(2n+1)) & \displaystyle  \frac{osp(M|2N)_k}{sp(2N)_k}  \\ \hline
	\displaystyle sp(2Mn) & \displaystyle  \frac{so(N+M)_k}{so(N)_k} \\ \hline
	\displaystyle sp(2m(2n+1)) & \displaystyle  \frac{sp(2N+2m)_k}{sp(2N)_k}  \\ \hline
	\displaystyle so(4mn) & \displaystyle  \frac{osp(N|2m)_k}{so(N)_k}\\ \hline
	\displaystyle osp(M (2 n+1)| 2 M n) & \displaystyle \frac{osp(M|2N)_k \oplus sp(2NM)_{-1/2}}{sp(2N)_{k-M/2}} \\ \hline
	\displaystyle osp(M (2 n -1)| 2M n) & \displaystyle \frac{so(N+M)_k\oplus so(NM)_1}{so(N)_{k+M}} \\ \hline
	\displaystyle osp(4 m n | 2m (2 n +1)) & \displaystyle  \frac{sp(2N+2m)_k \oplus so(4Nm)_1}{sp(2N)_{k+m}}  \\ \hline
	\displaystyle osp(4 m n | 2m (2 n -1)) & \displaystyle \frac{osp(N|2m)_k \oplus sp(2 N m )_{-1/2}}{so(N)_{k-2m}} \\ \hline
	\end{array} 
	\end{align*}
	\caption{Alternative proposals on the coset models whose symmetries are realized by the rectangular W-(super)algebras.}
	\label{fig:proposal2}
\end{table}

\subsection{Comparison with dual coset models}

We summarize the map of parameters for each dual coset model with a rectangular W-(super)algebra as its symmetry.

\subsubsection*{\underline{Type $so(M(2n+1))$}}

As the first example, we study the W-algebra with $so(M)$ symmetry obtained from the Hamiltonian reduction of $so(M(2n+1))$.
We propose that the symmetry can be realized by the coset \eqref{cosetsoII} along with \eqref{cosetsoI}.
The central charge of the coset is \eqref{centralsoII} and the level of  $so(M)$ affine symmetry is $-2 k$.
Compared with the central charge \eqref{csoI} and the level \eqref{lsoI} for the rectangular W-algebra,
we find that the correspondence happens at $\ell = - 2 k$ and $\lambda = 2 n+1$ with
\begin{align}
\lambda = \frac{k+1}{k+N+1} \, , \quad \lambda = - \frac{k+1}{k + N -M/2 + 1} \, .
\end{align}

\subsubsection*{\underline{Type $sp(2 M n)$}}

We move to the W-algebra with $so(M)$ symmetry obtained from the Hamiltonian reduction of $sp(2 n M)$.
The symmetry is proposed to be realized by the coset \eqref{cosetsoI} along with \eqref{cosetsoII}.
The central charge of the coset is \eqref{centralsoI} and the level of  $so(M)$ affine symmetry is $ k$.
Compared with  \eqref{csoII} and \eqref{lsoII},
the correspondence is realized with $\ell = k$ and $\lambda = 2 n$ with
\begin{align}
\lambda = \frac{k}{k+N-2} \, , \quad \lambda = - \frac{k}{k + N + M - 2} \, .
\end{align}

\subsubsection*{\underline{Type $sp(2 m (2 n +1))$}}

We then consider the W-algebra with $sp(2m)$ symmetry obtained from the Hamiltonian reduction of $sp(2 m (2 n +1))$.
The dual coset  is proposed as \eqref{cosetspII} in addition to \eqref{cosetspI}.
The central charge of the coset is \eqref{centralspII} and the level of  $sp(2m)$ affine symmetry is $ k$.
Compared with \eqref{cspI} and  \eqref{lspI},
we find the map of parameters as $\ell = k$ and $\lambda = 2 n+1$ with
\begin{align}
\lambda = \frac{k+1}{k+N+1} \, , \quad \lambda = - \frac{k+1}{k + N + m +1} \, .
\end{align}

\subsubsection*{\underline{Type $so(4 m n )$}}

As the final example of bosonic W-algebras, we examine the one with $sp(2m)$ symmetry from the Hamiltonian reduction of $so(4 m n)$.
We propose the dual coset  as \eqref{cosetspI} along with \eqref{cosetspII}.
The central charge of the coset is \eqref{centralspI} and the level of  $sp(2m)$ affine symmetry is $ -k/2$.
Comparison with \eqref{cspII} and  \eqref{lspII} leads to   $\ell = -k/2$ and $\lambda = 2 n$ with
\begin{align}
\lambda = \frac{k}{k+N-2} \, , \quad \lambda = - \frac{k}{k + N -2 m -2} \, .
\end{align}

\subsubsection*{\underline{Type $osp(M(2n+1)|2Mn)$}}

As the fist example of  $\mathcal{N}=1$ W-algebras, we study the one with $so(M)$ symmetry obtained from the Hamiltonian reduction of $osp(M(2n+1)|2Mn)$.
We propose the dual coset  as \eqref{dualcosetssoII} along with \eqref{dualcosetssoI}.
The central charge of the coset is \eqref{centralssoII} and the levels of  $so(M)$ affine symmetries are $- 2 k$ and $ - 2 N$.
Comparison with \eqref{cssoI} and  \eqref{l1l2ssoI},
we find that the correspondence happens at 
\begin{align}
\ell_1 = - 2 k \, , \quad \ell_2 = - 2 N \, , \quad \lambda \equiv \frac{ N}{ k+ N -M/2+1} = - 2 n
\end{align}
or
\begin{align}
\ell_1 = - 2 N \, , \quad \ell_2 = - 2 k \, , \quad \lambda \equiv \frac{ k}{ k+ N -M/2+1} = - 2 n \, .
\end{align}

\subsubsection*{\underline{Type $osp(M(2n-1)|2 M n)$}}

We move to the $\mathcal{N}=1$ W-algebra with $so(M)$ symmetry as  the Hamiltonian reduction of $osp(M(2n-1)|2 M n)$.
We propose that the symmetry is realized by the coset \eqref{dualcosetssoI} as well as \eqref{dualcosetssoII}.
The central charge of the coset is \eqref{centralssoI} and the levels of  $so(M)$ affine symmetries are $  k$ and $ N$.
Compared with \eqref{cssoII} and  \eqref{l1l2ssoII},
we find that the map of parameters as
\begin{align}
\ell_1 = k \, , \quad \ell_2 = N \, , \quad \lambda \equiv \frac{N-2}{k+N+M-2} = - 2 n +1
\end{align}
or
\begin{align}
\ell_1 =  N \, , \quad \ell_2 =  k \, , \quad \lambda \equiv \frac{k-2}{k+N+M-2} = - 2 n + 1\, .
\end{align}

\subsubsection*{\underline{Type $osp(4mn | 2 m (2 n +1))$}}

We then consider the $\mathcal{N}=1$ W-algebra with $sp(2m)$ symmetry as  the Hamiltonian reduction of $osp(4mn | 2 m (2 n +1))$.
The symmetry is proposed to be given by the coset \eqref{dualcosetsspII} as well as \eqref{dualcosetsspI}.
The central charge of the coset is \eqref{centralsspII} and the levels of  $sp(2m)$ affine symmetries are $  k$ and $ N$.
Comparison with \eqref{csspI} and  \eqref{l1l2sspI} leads to
\begin{align}
\ell_1 = k \, , \quad \ell_2 = N \, , \quad \lambda \equiv \frac{N}{k+N + m+1} = - 2 n 
\end{align}
or
\begin{align}
\ell_1 =  N \, , \quad \ell_2 =  k \, , \quad \lambda \equiv \frac{k}{k+N +m+1}= - 2 n \, .
\end{align}

\subsubsection*{\underline{Type $osp(4 m n | 2 m (2 n -1) )$}}

As the final example, we examine the $\mathcal{N}=1$ W-algebra with $sp(2m)$ symmetry obtained from  the Hamiltonian reduction of $osp(4 m n | 2 m (2 n -1) )$.
The symmetry is proposed to be given by the coset \eqref{dualcosetsspI} as well as \eqref{dualcosetsspII}.
The central charge of the coset is \eqref{centralsspI} and the levels of  $sp(2m)$ affine symmetries are $ - k/2$ and $ - N/2 $.
Compared with \eqref{csspII} and  \eqref{l1l2sspII}, we find a map as
\begin{align}
\ell_1 = - k /2 \, , \quad \ell_2 = - N /2 \, , \quad \lambda \equiv \frac{N-2}{k + N -2 m- 2} = - 2 n + 1
\end{align}
or
\begin{align}
\ell_1 =  - N /2 \, , \quad \ell_2 =  - k /2 \, , \quad \lambda \equiv \frac{k-2}{k+ N -2 m- 2}= - 2 n + 1\, .
\end{align}

\subsection{Symmetry of dual coset models}

We check the rectangular W-algebras with generators of spin up to two can be realized from the proposed cosets by explicitly constructing symmetry generators.

\subsubsection*{\underline{Type $sp(2 M n)$ with $n=1$}}

We consider the W-algebra with $so(M)$ symmetry obtained from the Hamiltonian reduction of $sp(2 M)$.
We construct symmetry generators in the coset \eqref{cosetsoI} and check that they generate the OPEs obtained in subsection \ref{sec:OPEs}. For the generators of $so(M+N)$, we use $(M+N)\times (M+N)$ matrices
\begin{align}
M_{M+N} = 
\begin{pmatrix}
A_M & B_{M,N} \\
C_{N,M} & D_N 
\end{pmatrix}  
\end{align}
with
\begin{align}
A^\text{tr} _M = - A_M \, , \quad D^\text{tr}_N = - D_N \, , \quad B^\text{tr}_{M,N} = - C_{N,M} \, . 
\end{align}
Denoting $t^a_M,t^p_N$ as the generators of $so(M),so(N)$, we express the generators of $so(M+N)$ as
\begin{align}
t^a_{M + N} =
\begin{pmatrix}
t^a_M & 0 \\
0 & 0
\end{pmatrix} \, , \quad
t^p_{M+N}
=
\begin{pmatrix}
0 & 0 \\
0 & t^p_{N} 
\end{pmatrix} \, , \quad
t^i_{M + N} =
\begin{pmatrix}
0 & t^i_{M,N} \\
- (t^i_{M,N})^\text{tr} & 0
\end{pmatrix} \, .
\end{align}
Here $t^i_{M,N}$ has one non-zero element, which is given by one.
We denote generators for the symmetric representation of $so(M)$ as $t^\alpha_M$ and introduce
\begin{align}
t^\alpha_{M + N} = 
\begin{pmatrix}
t^\alpha_M & 0 \\
0 & 0
\end{pmatrix} \, .
\end{align}
With $t^P_{M+N}=(t^a_{M+N},t^\alpha_{M+N},t^p_{M+N},t^i_{M+N})$, we introduce the metric and the invariant tensors as in \eqref{tPQ0} and \eqref{tPQ}. The OPEs of $so(M+N)$ currents are
\begin{align}
J^X (z) J^Y (0) \sim \frac{k/2 g^{XY}}{z^2} + \frac{i f^{XY}_{~~~Z} J^Z (0)}{z}
\end{align}
with $J^X = (J^a, J^p , J^i)$.
With these notations, the spin one generator for the symmetry algebra is given by $J^a$ with level $k$.
A spin two generator is the energy-momentum tensor, which can be constructed in the standard way \cite{Goddard:1984vk}. Other spin two generators can be found as
\begin{align}
\hat Q^\alpha = d^\alpha_{~ij} (J^i J^j) - \frac{N}{2 k + M - 4} d^\alpha_{~ab} (J^a J^b) \, .
\label{spin2soI}
\end{align}
We have checked that the operators satisfy the desired OPEs in subsection \ref{sec:OPEs} up to null vectors at $k = - (2N + 2M - 4)/3$ for $M=3,4,5,6$ and $N=2,3,4,5$

\subsubsection*{\underline{Type $so(4 m n )$ with $n=1$}}

We then examine the W-algebra with $sp(2m)$ symmetry obtained from the Hamiltonian reduction of $so(4 m)$. The OPEs among generators are examined  in subsection \ref{sec:OPEs}, and we reproduce them from 
symmetry generators in the coset \eqref{cosetspI}. 
For the generators of $osp(N|2m)$, we use $(2m+N)\times (2m+N)$ matrices
\begin{align}
M_{2m|N} = 
\begin{pmatrix}
A_{2m} & B_{2m,N} \\
C_{N,2m} & D_N 
\end{pmatrix}  
\end{align}
satisfying
\begin{align}
\Omega_{2m}^{-1} A^\text{tr} _{2m} \Omega_{2m} = - A_{2m} \, \quad D^\text{tr}_N = - D_N \, , \quad B^\text{tr}_{2m,N} \Omega_{2m}  =  C_{N,2m} \, , \quad \Omega_{2m}^\text{tr} = - \Omega_{2m} \, . 
\end{align}
Denoting $t^a_{2m},t^p_N$ as generators of $sp(2m),so(N)$, respectively, the generators of $osp(N|2m)$ are expressed as
\begin{align}
t^a_{2m|N} =
\begin{pmatrix}
t^a_{2m} & 0 \\
0 & 0
\end{pmatrix} \, , \quad
t^p_{2m|N}
=
\begin{pmatrix}
0 & 0 \\
0 & t^p_{N} 
\end{pmatrix} \, , \quad
t^i_{2m| N} =
\begin{pmatrix}
0 & t^i_{2m,N} \\
- (t^i_{2m,N})^\text{tr} \Omega_{2m} & 0
\end{pmatrix} \, .
\end{align}
Here $t^i_{2m,N}$ has one non-zero element, which is equal to one.
With $t^\alpha_{2m}$ as the generators of anti-symmetric representation of $sp(2m)$, we introduce
\begin{align}
t^\alpha_{2m| N} = 
\begin{pmatrix}
t^\alpha_{2m} & 0 \\
0 & 0
\end{pmatrix} \, .
\end{align}
With $t^P_{2m|N}=(t^a_{2m|N},t^\alpha_{2m|N},t^p_{2m|N},t^i_{2m|N})$,  the metric and the invariant tensors are as in \eqref{tPQ0} and \eqref{tPQ}. The $osp(N|2M)$ currents include bosonic ones $J^a,J^p$ and fermionic ones $\tilde J^i$. The non-trivial OPEs are 
\begin{align}
\begin{aligned}
&J^a (z) J^b (0) \sim \frac{- k/2  g^{ab} }{z^2} + \frac{i f^{ab}_{~~c} J^c (0)}{z} \, , \quad
J^p (z) J^q (0) \sim \frac{- k/2 g^{pq}}{z^2} + \frac{i f^{pq}_{~~~r} J^r (0)}{z} \, , \\
&\tilde J^i (z) \tilde J^j (0) \sim \frac{- k/2 g^{ij}}{z^2} + \frac{d^{ij}_{~~a} J^a (0)+ d^{ij}_{~~p} J^p (0)}{z} \, ,\\ & J^a (z) \tilde J^i (0) \sim \frac{i f^{ai}_{~~j} \tilde J^j (0)}{z} \, , \quad
J^p (z) \tilde J^i (0) \sim \frac{i f^{pi}_{~~j} \tilde J^j (0)}{z} \, .
\end{aligned} \label{ospOPE2}
\end{align}
The spin one generator of the symmetry algebra is given by $J^a$ with level $-k/2$.
A spin two generator is the energy-momentum tensor constructed by following \cite{Goddard:1984vk}. Other spin two generators can be found as
\begin{align}
\hat Q^\alpha = d^\alpha_{~ij} (\tilde J^i \tilde J^j) + \frac{N}{ 2 k - 2m  - 4} d^\alpha_{~ab} (J^a J^b) \, . \label{spin2spI}
\end{align}
We have checked that the operators satisfy the desired OPEs in subsection \ref{sec:OPEs} up to null vectors at $k= (- 2 N + 4m + 4)/3$ for $m=2,3,4$ and $N=2,3,4,5$.%
\footnote{Exceptions are with $(m,N)=(3,2),(4,4)$, where $\hat Q^\alpha$ become null vectors as $\hat Q^\alpha (z) \hat Q^\beta (0) \sim \mathcal{O}(z^{-3})$. With these values of $(m,N)$, the parameter $k$ becomes an integer $(k=4)$, and  coset type truncations would appear as discussed at the end of subsection \ref{sec:OPEs}.}

\subsubsection*{\underline{Type $osp(M(2n-1)|2 M n)$ with $n=1$}}

As an example of $\mathcal{N}=1$ W-algebras, we examine the one with $so(M)$ symmetry obtained from the Hamiltonian reduction of $osp(M|2M)$. The OPEs for the algebra were computed in subsection \ref{sec:superOPEs}, and we reproduce them from the symmetry generators of the coset \eqref{dualcosetssoI}. We use the same notation above for the $so(N+M)$ generators and currents.
In addition, there are free fermions $\psi^i$ from $so(NM)_1$ with
\begin{align}
\psi^i (z) \psi^j (0) \sim \frac{g^{ij}}{z} \, .
\end{align}
With them, we can construct $so(M)_N$ and $so(N)_M$ currents as
\begin{align}
J^a_f = - \frac{i}{2} f^{a}_{~ij} (\psi^i \psi^j) \, , \quad J^p_f = - \frac{i}{2} f^{p}_{~ij} (\psi^i \psi^j) \, .
\end{align}
The $so(N)_{k+M}$ currents in the coset \eqref{dualcosetssoI} are given by
\begin{align}
\hat J^p = J^p + J^p_f \, .
\end{align}
The spin one generators of the symmetry algebra are  $J^a$ and $K^a = J^a_f$.
For spin 3/2, we have
\begin{align}
G = g_{ij} (J^i \psi^j) \, , \quad G^a = - i f^{a}_{~ i j} (J^i \psi^j ) \, , \quad G^\alpha = d^{\alpha}_{~ij} (J^i \psi^j) \, .
\end{align}
A spin two current is the energy-momentum tensor and the others are the same as the bosonic case in \eqref{spin2soI}.
We have checked that the OPEs in subsection \ref{sec:superOPEs} are reproduced for several examples if we set
\begin{align}
\ell_1 = k \, , \quad \ell_2 = N \, , \quad k = - 2 N - M + 4 \, .
\end{align}

\subsubsection*{\underline{Type $so(4 m n | 2 m (2 n -1) )$ with $n=1$}}

We finally consider the $\mathcal{N}=1$ W-algebra with $sp(2m)$ symmetry obtained from the Hamiltonian reduction of $osp(4m |2m)$. We construct the symmetry generators of the coset \eqref{dualcosetsspI} and check the OPEs in subsection \ref{sec:superOPEs} are reproduced. 
We use the same notation above for the $osp(N|2M)$ generators and currents.
The coset also includes the symplectic bosons $\varphi^i$ from $sp(2 N m)_{-1/2}$ with
\begin{align}
\varphi^i (z) \varphi^j (0) \sim \frac{g^{ij}}{z} \, ,
\end{align}
and $sp(2m)_{-N/2}$ and $so(N)_{-2 m}$ currents can be constructed as
\begin{align}
J^a_f =  \frac{i}{2} f^{a}_{~ij} (\varphi^i \varphi^j) \, , \quad
 J^p_f =  \frac{i}{2} f^{p}_{~ij} (\varphi^i \varphi^j) \, .
\end{align}
The $so(N)_{k -2 m}$ currents in the coset \eqref{dualcosetsspI} are 
\begin{align}
\hat J^p = J^p + J^p_f \, .
\end{align}
The spin one generators of the symmetry algebra are  $J^a$ and $K^a = J^a_f$.
The spin 3/2 generators are
\begin{align}
G = g_{ij} (J^i \varphi^j) \, , \quad G^a =  i f^{a}_{~ i j} (J^i \varphi^j ) \, , \quad G^\alpha = - d^{\alpha}_{~ij} (J^i \varphi^j) \, .
\end{align}
A spin two current is the energy-momentum tensor and the others are the same as the bosonic case in \eqref{spin2spI}.
We have checked that the OPEs  in subsection \ref{sec:superOPEs} are reproduced for several examples if we set
\begin{align}
\ell_1 = -k/2 \, , \quad \ell_2 = -N/2 \, , \quad k = - 2 N + 2 m + 4 \, .
\end{align}


\providecommand{\href}[2]{#2}\begingroup\raggedright\endgroup

\end{document}